\documentclass[a4paper,11pt]{article}
\pdfoutput=1 

\usepackage{jheppub} 

\usepackage[T1]{fontenc} 

\usepackage[utf8]{inputenc}
\usepackage{tikz}
\usepackage{capt-of}

\usepackage{comment}

\addtocontents{toc}{\protect\enlargethispage{\baselineskip}}

\usepackage[rightcaption]{sidecap}
\usepackage{graphicx}
\graphicspath{ {./images/} }

\newcommand{\m}{\mu}
\newcommand{\n}{\nu}
\newcommand{\p}{\vec\nabla}

\newcommand{\vev}[1]{{\left< {#1} \right>}}

\newcommand{\cur}[1]{{\left( {#1} \right)}}
\newcommand{\squ}[1]{{\left[ {#1} \right]}}

\newcommand{\tr}{{\rm tr\,}}
\newcommand{\cA}{{\mathcal A}}
\newcommand{\cB}{{\mathcal B}}

\newcommand{\cD}{{\mathcal D}}
\newcommand{\cK}{{\mathcal K}}
\newcommand{\cL}{{\mathcal L}}

\newcommand{\cQ}{{\mathcal Q}}

\newcommand{\cM}{{\mathcal M}}
\newcommand{\cO}{{\mathcal O}}

\newcommand{\cF}{{\mathcal F}}
\newcommand{\cG}{{\mathcal G}}

\newcommand{\cH}{{\mathcal H}}

\title{\boldmath The coset construction for non-equilibrium systems}

\author[a]{Michael J. Landry}

\affiliation[a]{Department of Physics, Center for Theoretical Physics,\\ Columbia University, 538W 120th Street, New York, NY, 10027, USA}

\emailAdd{ml2999@columbia.edu}

\abstract{We propose a systematic coset construction of non-equilibrium effective field theories (EFTs)  governing the long-distance and late-time dynamics of relativistic, finite-temperature  condensed matter systems. Our non-equilibrium coset construction makes significant advances beyond more standard coset constructions in that it takes advantage of recently-developed techniques, which allow the  formulation of non-equilibrium effective actions that account for quantum and thermal fluctuations as well as  dissipation. Because these systems exist at finite temperature, the EFTs live on the closed-time-path of the Schwinger-Keldysh contour. Since the coset construction and the non-equilibrium effective actions may be unfamiliar to many readers, we include brief introductions to these topics in an effort to make this paper self-contained. To demonstrate the legitimacy of this coset construction, we successfully reproduce the known EFTs for fluids and superfluids at finite temperature. Then, to demonstrate its utility, we construct novel EFTs for solids, supersolids, and four phases of liquid crystals, all at finite temperature.  We thereby combine the non-equilibrium effective action and the coset construction to create a powerful tool that can be used to study many-body systems out of thermal equilibrium.}

\arxivnumber{1912.12301}

\begin{document} 
\maketitle
\flushbottom

\section{Introduction}

When trying to understand a many-body system, the goal is often not to keep track of all of the degrees of freedom; there are simply too many. Instead, the aim is to focus only on quantities that persist over long distances and extended time scales. To do this, one coarse-grains over the microscopic, or ultraviolet (UV) degrees of freedom to obtain an effective theory of the macroscopic, or infrared (IR) observables. Often, one of the most important guiding principles in constructing such effective theories are symmetries and their corresponding Noether currents. For example the standard formulation of hydrodynamics comes from the equations for conservation of energy, momentum, and particle number~\cite{FD 1}. Even though such an approach is heavily based in symmetry considerations it is not very systematic as other non-symmetry constraints must be imposed. In particular, it relies on the somewhat ill-defined notion of `near-equilibrium' to construct constitutive relations among the conserved quantities and local thermodynamic parameters.\footnote{Near-equilibrium thermodynamics in the usual formulation is ill-defined in the sense that thermodynamic parameters like temperature and chemical potential are only rigorously defined in true equilibrium.} Once this is done, the second law of thermodynamics must be imposed by hand, putting constraints on various coefficients. While this method has led to very powerful and general statements about hydrodynamic systems, there remain some difficulties. In particular, it is not at all clear what are the rules of the game for constructing hydrodynamic theories. For example, one might wonder if there are additional constraints beyond the second law of thermodynamics that one must impose. Additionally, a way to extend hydrodynamic theories to methodically account for thermal fluctuations or quantum effects is not at all clear.  An approach that utilizes symmetries as the {\it only} input could resolve some of these issues. 

In order to make the approach to hydrodynamics---and many-body physics as a whole---more systematic, over the last decade or so there has been a great effort to reformulate hydrodynamics from the point of view of an effective action~\cite{H. Liu,H. Liu 2,H. Liu 3,H. Liu 2.2,H. Liu 2.3,H. Liu 2.1,H. Liu 4,Nicolis,Zoology,More gapped Goldstones,coset,Finite T superfluid,Harder:2015nxa,Constraints on Fluid,Jensen:2017kzi,Jensen:2018hse,Jensen:2018hhx,Towards hydrodynamics,Kovtun:2014hpa,Grozdanov:2013dba,Haehl:2017zac,Haehl:2018lcu,Fluid manifesto,Hongo 1,Hongo 2,Hongo 3,Hongo 4,Volkov}. The reasons for doing so are as follows. Effective actions provide powerful tools for describing systems in high-energy and particle physics because they can be constructed from symmetry principles alone. Once the symmetries and field content are specified, one constructs the effective action by writing down a linear combination of all terms that are consistent with the symmetries. 
Thus, there is no guesswork when formulating effective actions. One can then express local thermodynamic quantities like temperature and chemical potential in terms of the fields, yielding a more rigorous understanding of the concept of near-equilibrium thermodynamics. Applying these methods to condensed matter systems has provided powerful tools of analysis.

From the effective field theory (EFT) perspective, many-body systems can often be understood as systems that spontaneously break spacetime symmetries. In fact, it is often possible to classify the state of matter of a many-body system by its spontaneous symmetry breaking (SSB) pattern alone~\cite{Zoology}. Then, the long-distance and late-time dynamics are described entirely by Goldstone bosons. For our purposes, this is very good news: there is a powerful and systematic way to construct EFTs of Goldstones known as the coset construction~\cite{Ogievetsky,Ivanov and Ogievetsky,Low,Weinberg,Wheel}.  This construction takes the symmetry-breaking pattern---which is specified by the state of matter of the system---as the only input. The coset construction has been used to formulate numerous EFTs describing various states of matter~\cite{coset}. 

Historically, the EFT approach to many-body physics has been unable to account for statistical fluctuations and dissipation. The reason is that ordinary actions describe noiseless, conservative systems and are therefore incapable of accounting for stochastic and dissipative dynamics. However, this changed with recent work~\cite{H. Liu,H. Liu 2, H. Liu 3,H. Liu 2.2,H. Liu 2.3,H. Liu 2.1,H. Liu 4} and~\cite{Towards hydrodynamics,Kovtun:2014hpa,Grozdanov:2013dba,Haehl:2017zac,Haehl:2018lcu,Fluid manifesto} in which a new kind of non-equilibrium action was formulated using the in-in formalism on the Schwinger-Keldysh (SK) contour. Using symmetry principles alone these new non-equilibrium EFTs can account for both statistical fluctuations as well as dissipation. Because these EFTs are constructed on the SK contour, the field content is doubled. In addition to the global symmetry group, there exist emergent gauge symmetries at finite temperature, the origins of which remain poorly understood~\cite{H. Liu}. 

Nevertheless, despite significant advances, in the current literature there exist very few systems for which the non-equilibrium EFTs are known. The reason is that no overarching understanding of the field content,  emergent gauge symmetries, or symmetry-breaking pattern has yet been provided for non-equilibrium systems. In this paper, we address these problems by combining two powerful tools: the non-equilibrium EFT approach to many-body systems and the coset construction.\footnote{The coset construction was recently used in~\cite{Hongo 1} to build simple EFTs on the SK contour; however their approach only dealt with internal symmetries.} We term the result the {\it non-equilibrium coset construction}. 

Our approach is as follows: We start from the observation that at finite temperature, there exist Goldstone-like excitations corresponding to each symmetry generator even if it is not spontaneously broken. However, the Goldstones corresponding to the unbroken generators behave very differently than those arising from SSB. In particular the Goldstones corresponding to unbroken generators have infinitely many gauge symmetries analogous to the chemical shift symmetries of~\cite{Nicolis}. These gauge symmetries lead to diffusive behavior. Since the non-equilibrium effective action is defined on the SK contour, the field content is doubled. This requires the introduction of two cosets; one for each leg of the SK contour. We use these cosets to construct building-blocks for the non-equilibrium effective actions that transform covariantly under both the global symmetries  as well as the chemical shift-type gauge symmetries.

To demonstrate the validity of this new coset construction, we use it to formulate actions for already known EFTs, namely those of fluids and superfluids at finite temperature. Then, to demonstrate the utility of our approach, we construct novel EFTs for solids, supersolids, and four phases of liquid crystals, all at finite temperature. We thus vastly expand the number of known non-equilibrium EFTs. Moreover, our formalism can be used to construct EFTs for essentially any state of matter that is describable by  Goldstone (or Goldstone-like) excitations. The coset construction and the non-equilibrium EFT formalism may be unfamiliar to some readers. To make this paper as self-contained as possible, sections \S\ref{coset review} and \S\ref{non-eq review} give quick reviews of these topics. Readers already familiar with these topics can skip these sections entirely. 

Throughout this paper we will use the `mostly plus convention' so the Minkowski metric takes the form $\eta_{\m\n}=\text{diag}(-,+,+,+)$.

\section{The zero-temperature coset construction: a review}\label{coset review}

Consider a relativistic quantum field theory whose full symmetry group is $\cG$, which includes both internal symmetries as well as the Poincaré group. Suppose that the ground state of the system spontaneously breaks the symmetry group $\cG$ to the subgroup $\cH$. Then the IR dynamics of the system are described by Goldstone modes. If only internal symmetries are spontaneously broken, then for every broken symmetry generator in the coset $\cG/\cH$, there is a corresponding Goldstone boson. However, if spacetime symmetries are spontaneously broken, there are often fewer Goldstones than broken symmetry generators~\cite{Ogievetsky,Ivanov and Ogievetsky,Low}. 

Often, the only gapless modes in the system are the Goldstones. If we are only concerned with the deep IR dynamics, we can integrate out all non-Goldstone modes to obtain an effective action for the Goldstones. It turns out that there is a systematic method for writing down  Goldstone effective actions known as the coset construction. This section will give a brief review of the coset construction for spontaneously broken internal and spacetime symmetries. For in-depth discussions of the coset construction for spontaneously broken internal symmetries, consult~\cite{Weinberg} and for spontaneously broken spacetime symmetries, consult~\cite{Wheel}.

The coset construction is an especially powerful technique since it provides a systematic method for generating effective actions of Goldstone bosons using the SSB pattern as the only input. Supposing the symmetry-breaking patten is $\cG\to\cH$, we represent the symmetry generators  by
\begin{equation}\begin{split} \bar P_\mu & =\text{unbroken translations},
\\ T_A & = \text{other unbroken generators},
\\ \tau_\alpha & =\text{broken generators}, 
 \end{split}\end{equation}
 where the generators $\tau_\alpha$ and $T_A$ may be some combination of internal and spacetime generators and  we have assumed that there exist some notions of spacetime translations that remain unbroken.  In this way, states can still be classified according to the corresponding notions of energy and momentum~\cite{Zoology}. Importantly, we do not require that the unbroken generators $\bar P_\mu$ be the original Poincaré translation generators (represented by $P_\m$); instead they can be some linear combination of $P_\m$ and internal symmetry generators~\cite{Zoology}. It will turn out that although $\bar P_\m$ and $T_A$ both refer to unbroken generators, they will play very different roles in the coset construction. Therefore, it is convenient to define the subgroup $\cH_0\subset \cH$ that is generated exclusively by $T_A$.  

The EFT of the Goldstones must be invariant under the full symmetry group $\cG$. While the action of the unbroken symmetry subgroup $\cH$ is linearly realized on the fields, the action of the broken coset $\cG/\cH$ is nonlinearly realized. In general, therefore, generic symmetry transformations will act in a highly non-trivial manner on the Goldstones. As a result writing down the most general effective action consistent with symmetries can be rather challenging unless we use the construction that follows.

We are interested in ensuring that the effective action remain invariant under all symmetry transformations. The symmetries that act linearly are easy to deal with, but those that act non-linearly present more of a challenge. By definition, spontaneously broken generators act non-linearly on the fields while unbroken generators act linearly. However, unbroken translations act non-linearly on the {\it coordinates}, meaning that the coset of symmetry generators that have some sort of non-linear action is $\cG/\cH_0$.\footnote{This is a somewhat hand-wavy justification for including $e^{ix^\m\bar P_\m}$ in the coset. But in any case it turns out that~(\ref{ordinary coset}) has the correct symmetry properties for our purposes.} It is convenient to parameterize this coset by
\begin{equation}\label{ordinary coset} \gamma( x,\pi] = e^{i x^\mu \bar P_\m} e^{i \pi^\alpha(x) \tau _\alpha}.\end{equation} 
Then, up to normalization, $\pi^\alpha(x)$ correspond to the Goldstones. 

By the definition of the coset $\cG/\cH_0$, it is possible to express any element of $\cG$ as the product of an element of the coset and an element of $\cH_0$.  As a result, for any element $g\in\cG$, we may write
\begin{equation}\label{coset trans} g \cdot \gamma(x,\pi] = \gamma(x',\pi'] \cdot h_0(x,\pi,g], \end{equation}
where $h_0$ is some element of $\cH_0$ and $x'$ and $\pi'$ are transformed coordinates and fields. It is important to note that for given $\pi$ and $g$, the terms on the r.h.s. of~(\ref{coset trans}) can be explicitly computed if we know the commutation relations among the generators. We can therefore read off the transformations of the coordinates and Goldstones under an arbitrary symmetry transformation from~(\ref{coset trans}).  In particular,  under the transformation by $g$, we have $x\to x'$ and $\pi\to\pi'$. 

Using the parameterization~(\ref{ordinary coset}), we can construct the Maurer-Cartan one-form  $g^{-1} d g$. This one-form has the property that it can always be expressed as a linear combination of the symmetry generators~\cite{Weinberg,Wheel}, so we have
\begin{equation} g^{-1}\partial_\m g = i E_\m^\n \cur{\bar P_\n +\nabla_\n \pi^\alpha \tau_\alpha+\cB_\n^A T_A}.  \end{equation}
Using the symmetry algebra alone, it is possible to compute the coefficients of each generator in the above expression. 
It can be checked that $\nabla_\m \pi^\alpha$ transforms covariantly under the full symmetry group $\cG$; we therefore will refer to it as the covariant derivative of $\pi$. Additionally, $\cB_\n^A$ transforms like a connection (i.e. gauge field) and we can therefore use it to compute higher-order covariant derivatives by
\begin{equation} \label{covariant derivative coset} \nabla_\m^{\cH} = (E^{-1})^\n_\m \partial_\n + i \cB_\m^A T_A .\end{equation}
And finally, $E_\m^\n$ serves as a vierbein; in particular, the invariant integration measure is $d^4 x \det E$. It should be noted that the indices $\mu,\nu$ need not be contracted in the usual way if the Lorentz group is broken. 

After computing $g^{-1}\partial_\m g$, it is then easy to identify the covariant building-blocks. At leading order in the derivative expansion we have $\nabla_\m \pi^\alpha$; and higher-order-derivative terms  are given by 
$\nabla^\cH_\m \cur{\nabla_\n \pi^\alpha},$ 
$\nabla^\cH_\m \nabla_\n^\cH \cur{\nabla_\rho \pi^\alpha},$ etc. Then, the symmetry-invariant terms of the effective action are simply constructed by taking manifestly $\cH$-invariant combinations of these covariant terms. If boosts are broken but rotations are not, then examples of such invariant terms are $(\nabla_0\pi^\alpha)^2$ and $(\nabla_i\pi^\alpha)^2$, where repeated indices are summed over.

\subsection{Inverse Higgs}

We mentioned earlier that when only internal symmetries are broken, the number of Goldstones exactly matches the number of broken generators. However, when spacetime symmetries are broken, this need not be the case. In this section, we will see how this works from the perspective of the coset construction. 

Pragmatically, the rules of the game are as follows:
Suppose that the commutator between an unbroken translation generator $\bar P$ and a broken generator $\tau '$ contains another broken generator $\tau$, that is $[\bar P, \tau '] \supset \tau$. Suppose further that  $\tau$ and $\tau '$ do not belong to the same irreducible multiplet under $\cH_0$. Then it turns out that it is consistent with symmetry transformations to set the covariant derivative of the $\tau$-Goldstone in the direction of $\bar P$ to zero. This gives a constraint that relates the $\tau '$-Goldstone to derivatives of the $\tau$-Goldstone, allowing the removal of the $\tau'$-Goldstone. The setting of this covariant derivative to zero is known as an inverse Higgs constraint.

The physical reasons for imposing these inverse Higgs constraints have been investigated in \cite{More gapped Goldstones,Low,UV completion without symmetry restoration}. Essentially, there are two possibilities. Firstly, sometimes when spacetime symmetries are spontaneously broken, the resulting Goldstones do not correspond to independent fluctuations. As a result, there can be multiple Goldstone field configurations that all correspond to the same physical state. From this viewpoint, the inverse Higgs constraints can be understood as a convenient choice of gauge-fixing condition.  Secondly, when a Goldstone can be removed via inverse Higgs constraints, it is often the case that if we did include the Goldstone in the effective action, it would have an energy gap. But we are often only interested in gapless excitations, so we may integrate out these fields. From this perspective, the inverse Higgs constraints correspond to integrating out gapped Goldstones.
For the purposes of this paper, with the exception of \S\ref{Fluids}, we will take a purely pragmatic approach and always impose inverse Higgs constraints whenever possible, remaining agnostic about the precise reasons for doing so.

\subsection{Zero-temperature superfluids: a simple example}

Thus far, our discussion of the coset construction has been rather abstract. This section will focus on a simple concrete example: the coset construction of the zero-temperature superfluid EFT. First, since our theory is relativistic, it ought to be Poincaré-invariant. In our `mostly plus' convention, the Poincaré algebra is 
\begin{equation}\begin{split}\label{Poincare}
i [J_{\m\n},J_{\rho\sigma}] &= \eta_{\n\rho} J_{\m\sigma} -\eta_{\m\rho}J_{\n\sigma} -\eta_{\sigma\m} J_{\rho\n} +\eta_{\sigma\n} J_{\rho\m},
\\ i[P_\m, J_{\rho\sigma}] & = \eta_{\m\rho} P_\sigma-\eta_{\mu\sigma} P_\rho,
\\ i[P_\m,P_\n] & = 0, 
\end{split}\end{equation}
where $P_\m$ are the translation generators and $J_{\m\n}$ are the Lorentz generators. 
From the EFT perspective, a superfluid is defined as a system that has a conserved $U(1)$ charge $Q$ such that both $Q$ and $P_0$ (i.e. time translation) are spontaneously broken but a diagonal subgroup, $\bar P_0\equiv P_0+\mu Q$ is preserved~\cite{Witten,Son}. Physically, $Q$ is the charge associated with particle-number conservation and $\mu$ is the chemical potential. Additionally, since every condensed matter system (including superfluids) has a zero-momentum frame, boosts are necessarily broken. As a result, the broken generators are the $U(1)$ charge $Q$ and Lorentz boots $K_i\equiv J_{0i}$. The unbroken translations are $\bar P_0=P_0+\mu Q$ and $\bar P_i=P_i$ for $i=1,2,3$ and the remaining unbroken generators are the spatial rotation generators $J_i\equiv \frac{1}{2} \epsilon^{ijk} J_{jk}$. 
With this symmetry-breaking pattern, the coset is then
\begin{equation} g = e^{i x^\m \bar P_\m} e^{i \pi(x) Q} e^{i \eta^i(x) K_i}. \end{equation}
By explicit computation, the Maurer-Cartan form is
\begin{equation} g^{-1}\partial_\m g= i E_\m ^\n \cur{ \bar P_\n + \nabla _\n \pi Q +\nabla_\n \eta^i K_i+\Omega_\n^i J_i}, \end{equation}
where 
\begin{equation} \begin{split}
E_\m^\n & = {\Lambda_\m}^\n,
\\ \nabla_\m \pi & = (E^{-1})^\n_\m \partial_\n \psi-\m \delta_\m^0,
\\ \nabla_\m \eta^i & =  (E^{-1})_\m^\n [\Lambda^{-1}\partial_\n \Lambda ]^{0i},
\\ \Omega_\m^i & = \frac{1}{2}(E^{-1})_\m^\n \epsilon^{ijk} [\Lambda^{-1}\partial_\n \Lambda]^{jk},
\end{split} \end{equation}
where 
${\Lambda^\m}_\n\equiv[e^{i\eta^i K_i} {]^\m}_\n$ and $\psi\equiv \m t+\pi$. Here, $\epsilon^{ijk}\Omega_\m^k$ can be thought of as the spin-connection (or at least the spatial components of it) and transforms as a gauge field under rotations. 

Now, notice that the commutator between unbroken spatial translations and broken boosts yields $[\bar P_i,K_j]\supset i \delta_{ij} \mu Q$, meaning that we may impose inverse Higgs constraints to remove the boost Goldstones. In particular, we may set to zero the covariant derivative of $\pi$ along the $i=1,2,3$ directions, yielding
\begin{equation} 0 = \nabla_i\pi ={ \Lambda^\m}_i \partial_\m \psi.  \end{equation}
These constraints can be solved to give a relation between the boost Goldstones and derivatives of the $U(1)$ Goldstones
\begin{equation}\label{IH example} \frac{\eta^i}{\eta} \tanh \eta = -\frac{\partial_i\psi}{\partial_0\psi}.  \end{equation}
Using these relations, we can remove the boost Goldstones as dynamical degrees of freedom. Thus, at leading order in the derivative expansion, the only covariant building-block is $\nabla_0\pi$. By plugging~(\ref{IH example}) into the expression for $\nabla_0\pi$, we find that
\begin{equation}\nabla_0\pi = \sqrt{- \partial_\m \psi\partial^\m \psi} -\m.  \end{equation}
Since $\mu$ is just a constant, we find that at leading order in derivatives, the zero-temperature EFT must be built out of an arbitrary function of 
\begin{equation}\label{zero T y} y \equiv  \sqrt{- \partial_\m \psi\partial^\m \psi}. \end{equation}
Since $\det E=1$, we have that the invariant integration measure is just $d^4 x$, so the leading-order action is
\begin{equation}\label{zero T superfluid EFT} S = \int d^4 x ~ P(y), \end{equation}
where $P$ is some function that is determined by the equation of state. Terms that are higher order in the derivative expansion can be constructed using $\nabla_\m\eta^i$ as well as $\Omega_\m^i$, which plays the role of a connection and can be used to create covariant derivatives of the form~(\ref{covariant derivative coset}). But to keep things as streamlined as possible, in this paper we will only construct leading-order actions.

\section{Non-equilibrium effective actions: a review}\label{non-eq review}

At zero temperature, the equilibrium state is described by a pure state, namely  the vacuum. At finite temperature, however, no such pure equilibrium state exists. Instead, the equilibrium state must be described by a mixed-state thermal density matrix of the form
\begin{equation}\rho=\frac{e^{-\beta_0 \bar P^0}}{Z},~~~~~~~~~~ Z=\tr e^{-\beta_0 \bar P^0}, \end{equation}
where $\beta_0$ is the inverse equilibrium temperature and $\bar P^0$ is the unbroken time-translation generator. In order to describe correlation functions in time, we must use the so-called in-in or Schwinger-Keldysh formalism~\cite{SK ref}. In this formalism, the sources are doubled. Letting $U(t,t',J)$ be the time-evolution operator from time $t'$ to $t$ in the presence of source $J$ for some field $\Psi$, the generating functional is
\begin{equation}\begin{split}e^{W[J_1,J_2]}&\equiv\label{SK}\tr{\squ{U(+\infty,-\infty;J_1)  \rho U^\dagger(+\infty,-\infty;J_2)}} 
\\ &\equiv\int _\rho\cD\Psi_1\cD\Psi_2e^{iS[\Psi_1,J_1]-iS[\Psi_2,J_2]},\end{split}\end{equation}
where in the path integral representation, we require that in the distant future, $\Psi_1({t\to\infty})=\Psi_2({t\to\infty})$, and the subscript $\rho$ indicates that field configurations are weighted by the thermal density matrix functional in the infinite past. Because there are two time-evolution operators such that one evolves forward in time and the other backward in time, we can conceive of the SK path integral as existing on a closed contour in time that starts at $t=-\infty$ goes to $t=+\infty$ and then returns again to $t=-\infty$. This contour is often referred to as a closed time path (CTP). 

It is possible to use this formalism to construct an effective action of the IR degrees of freedom on the CTP, known as the non-equilibrium effective action. Since the generating functional depends on two copies of the fields, the non-equilibrium effective action will have doubled field content.

Consider the path-integral representation of the generating functional~(\ref{SK}). We are interested in the meaning of the low-frequency, long-wavelength dynamics of the system. In typical Wilsonian fashion, we integrate out the fast modes to obtain an effective action for the slow modes. Suppose the fields can be divided up into IR and UV fields by $\Psi=\{\psi^{ir},\psi^{uv}\}$, where $\psi^{ir}$ represent the IR degrees of freedom and $\psi^{uv}$ represent the UV degrees of freedom. Define the effective action for the IR fields by\footnote{Here as well as in the remainder of the paper, we use $S$ to denote an ordinary action and $I$ to denote an action defined on the SK contour. }
\begin{equation}e^{i I_\text{EFT}[\psi^{ir}_1,\psi^{ir}_2;J_1,J_2]}=\int_\rho \cD\psi^{uv}_1\cD\psi^{uv}_2 e^{i S[\psi^{uv}_1,\psi^{ir}_1;J_1]-iS[\psi^{uv}_2,\psi^{ir}_2;J_2]},\end{equation}
then $e^{W[J_1,J_2]}=\int \cD[\psi^{ir}_1,\psi^{ir}_2]e^{iI_\text{EFT}[\psi^{ir}_1,\psi^{ir}_2;J_1,J_2]}$. 
Notice that the information contained in the density matrix $\rho$ is absorbed into the coefficients of $I_\text{EFT}$ so we do not need to include a density matrix for the IR fields $\psi^{ir}_{1,2}$. Finally we require that the Green functions of $I_\text{EFT}$ are path-ordered on the SK contour; the path-ordered Green functions are given in \cite{H. Liu}.

\subsection{Rules for constructing non-equilibrium EFTs}

Following the usual EFT philosophy, it is our hope that all of the complicated UV dynamics and information about $\rho$ can be absorbed in the low-frequency limit by a finite number of parameters in $I_\text{EFT}$ at any given order in the derivative and field expansions. It has been demonstrated that this is in fact the case~\cite{H. Liu}, but there are several important constraints that must be imposed upon $I_\text{EFT}$. We outline some important features of this effective action below without proof~\cite{H. Liu}.

\begin{itemize}
\item The UV action describing the system of interest is factorized by $S[\Psi_1;J_1]-S[\Psi_2;J_2]. $ The effective action, however, does not admit a factorized form. In general, there will exist terms that couple 1-and 2-fields in $I_\text{EFT}$. 

\item Notice that, while the coefficients of $S[\Psi_1;J_1]-S[\Psi_2;J_2]$ are purely real, the coefficients of $I_\text{EFT}[\psi^{ir}_1,J_1;\psi^{ir}_2,J_2]$ may be complex. There are three important constraints that come from unitarity, namely 
\begin{equation}\begin{split}\label{unitarity}I^*_\text{EFT}[\psi^{ir}_1,\psi^{ir}_2;J_1,J_2]&=-I_\text{EFT}[\psi^{ir}_2,\psi^{ir}_1;J_2,J_1]
\\  \text{Im} I_\text{EFT}[\psi^{ir}_1,\psi^{ir}_2;J_1,J_2]&\geq 0, ~~\text{for any} ~~\psi^{ir}_{1,2}, J_{1,2}
\\ I_\text{EFT}[\psi^{ir}_1=\psi^{ir}_2;J_1=J_2]&=0.
\end{split}\end{equation}
These conditions help to ensure that Green functions are path-ordered. 

\item Any symmetry of the UV action $S$ is a symmetry of $I_\text{EFT}$, except for time-reversing symmetries. The fact that these time-reversing transformations are not symmetries of the effective action allows the production of entropy. Because the field values on the 1-and 2-contours must be equal in the distant future, $\psi_1^{ir}$ and $\psi_2^{ir}$ must transform simultaneously under any global symmetry transformation. Thus, there is just one copy of the global symmetry group. 

\item If the equilibrium density matrix $\rho$ takes the form of a thermal matrix, $\rho\propto e^{-\beta_0 \bar P_0}$, then the partition function $W[J_1,J_2]$ obeys what are known as the KMS conditions. These KMS conditions for the partition function can be used to derive the so-called dynamical KMS symmetries of the effective action. The way these symmetries act is as follows. Suppose that the UV theory possesses some kind of anti-unitary time-reversing symmetry $\Theta$; at a minimum, the UV theory will be invariant under a simultaneous charge, parity, and time inversion. Then, setting the sources to zero, the dynamical KMS symmetries act on the fields by  
\begin{equation}\begin{split}\label{quantum dynamical KMS} \psi^{ir}_1(x)&\to\Theta\psi^{ir}_1(t-i\theta,\vec x)
\\ \psi^{ir}_2(x)&\to\Theta \psi^{ir}_2(t+i(\beta_0-\theta),\vec x),
\end{split}\end{equation} 
for any $\theta\in[0,\beta_0]$. It can be checked that these transformations are their own inverse, meaning that the dynamical KMS symmetries are discrete $\mathbb Z_2$ symmetries. These symmetries  involve temporal translations along the imaginary-time directions because the thermal density matrix can be interpreted as a time-translation operator by imaginary time $-i\beta_0$.  
As a result, they are non-local transformations. This non-locality may seem rather odd, but at any given order in the derivative expansion, these symmetries become local; one need only perform a Taylor series in $\theta$ and $\beta_0-\theta$.  Further, in the classical limit they become exactly local. To take the classical limit, it is convenient to perform a change of field basis by
\begin{equation}\label{r a variables} \psi^{ir}_r\equiv \frac{1}{2}\cur{\psi^{ir}_1+\psi^{ir}_2},~~~~~~~~~~\psi^{ir}_a\equiv\psi^{ir}_1-\psi^{ir}_2. \end{equation}  
Then the classical dynamical KMS symmetry transformations are 
\begin{equation}\begin{split}\psi^{ir}_r(x)&\to\Theta\psi^{ir}_r(x)
\\ \psi^{ir}_a(x)&\to\Theta\psi^{ir}_a(x)+i\Theta\squ{\beta_0\partial_t\psi^{ir}_r(x)}.
\end{split}\end{equation}
Notice that the change in $\psi^{ir}_a$ is proportional to the derivative of $\psi^{ir}_r$. Thus, when writing down terms of the effective action in the derivative expansion, it is natural to consider $\psi^{ir}_a$ and $\partial_t\psi^{ir}_r$ as contributing to the same order. 
\end{itemize} 

\subsection{The fluid worldvolume}

At finite temperature the effective action is often not defined on the physical spacetime. Instead, we must introduce the notion of the so-called fluid worldvolume. To see why this is so, we will reproduce the derivation of the fluid action presented in~\cite{H. Liu}. Consider a fluid with no conserved currents other than the stress-energy tensor. The sources for the stress-energy tensor are the metric tensors $g_{s\m\n}$, where $s=1,2$ indicates on which leg of the SK contour the metrics live. Then, the SK generating functional takes the form
\begin{equation} e^{W[g_{1\m\n},g_{2\m\n}] } =\text{tr}  \squ{ U(+\infty,-\infty; g_{1\m\n}) \rho U^\dagger(+\infty,-\infty; g_{2\m\n})},  \end{equation} 
where $U(+\infty,-\infty; g_{s\m\n})$ is the time-evolution operator in the presence of source $g_{s\m\n}$. 
Since the stress-energy tensor is conserved, the generating functional $W[g_{1\m\n},g_{2\m\n}] $ must be invariant under two independent diffeomorphism transformations. Let $\zeta^\m_s(x)$ for $s=1,2$ represent two different diffeomorphisms. Then we have
\begin{equation}W[g_{1\m\n},g_{2\m\n}] = W[g^{\zeta_1}_{1MN},g^{\zeta_2}_{2MN}] ,\end{equation} 
where $g_{sMN}^{\zeta_s}$ for $M,N=0,1,2,3$ denote diffeomorphism transformations of $g_{s\m\n}$. More explicitly, we have $g^{\zeta_s}_{sMN}(\phi) \equiv \frac{\partial\zeta_s^\m}{\partial\phi^M} g_{s\m\n}(\zeta_s(\phi)) \frac{\partial\zeta^\n_s}{\partial\phi^N}$. Now, we can use the Stückelberg trick and promote the gauge transformations to dynamical fields. In particular, we `integrate in' the fields $X_s^\m(\phi)$ such that the generating functional becomes  
\begin{equation}e^{W[g_{1\m\n},g_{2\m\n}] } = \int \cD X_{1}\cD X_2 ~e^{i I_\text{EFT}[G_{1MN},G_{2MN}]},\end{equation}
where 
\begin{equation} G_{sMN} \equiv  \frac{\partial X_s^\m}{\partial\phi^M} g_{s\m\n}(X_s(\phi)) \frac{\partial X^\n_s}{\partial\phi^N} \end{equation}
are  pull-back metrics. 
Notice that when performing the Stückelberg trick, we had to introduce the coordinates $\phi^M$ for $M=0,1,2,3$. We will refer to $\phi^M$ as `fluid coordinates' and the corresponding manifold on which they live as the `fluid worldvolume.' Then the fields $X_s^\m(\phi)$ are the dynamical fields and describe the embedding of the fluid worldvolume into the physical spacetime. In a certain sense, it is always possible to `integrate in' these Stückelberg fields as long as $W[g_{1\m\n},g_{2\m\n}]$ is diffeomorphism invariant. However, there is no guarantee that the resulting EFT will be non-trivial; for example, if the equilibrium density matrix is a pure, zero-temperature ground state without any SSB then the fields $X_s^\m$ are pure gauge and hence have no dynamics. It turns out that to ensure the system exists in fluid phase, we must require that the fluid coordinates enjoy partial diffeomorphism symmetries:
\begin{itemize}
\item Fluid elements are indistinguishable from one another, so we should be able to freely relabel them in a time-independent manner. We therefore require invariance under
\begin{equation}\label{spat diff fl}
\phi^I\to{\phi'}^I(\phi^J),
\end{equation}
for $I,J=1,2,3$. 
\item Treating each volume element as a point particle, we require independent world-line reparametrization invariance,
\begin{equation}
\phi^0\to{\phi'}^0=f(\phi^0,\phi^I),~~~~~~~~~~\tau(\phi^0,\phi^I) \to \frac{\partial f}{\partial \phi^0}\tau ({\phi'}^0,\phi^I),
\end{equation}
where $\tau(\phi^0,\phi^I)$ plays the role of the worldline einbein for the fluid volume element at~$\phi^I$. We can then gauge-fix $\tau=1$, in which case we have the residual symmetry
\begin{equation}\label{time diff fl}
\phi^0\to{\phi'}^0(\phi^I),
\end{equation}
for $I=1,2,3$. 
\end{itemize}

The partial diffeomorphism symmetries~(\ref{spat diff fl}) and~(\ref{time diff fl}) can be succinctly encapsulated by the transformation law
\begin{equation}\phi^M\to\phi^M+\xi^M(\phi^I),\end{equation}
where $\xi^M(\phi^I)$ are arbitrary functions of the {\it spatial} fluid coordinates $\phi^I$ for $I=1,2,3$. The justifications provided above for imposing these diffeomorphism symmetries are merely included to build the reader's intuition; it is not understood from first principles how they arise, only that they are necessary to describe systems in fluid phase. Interested readers can consult Appendix \ref{Symmetry origin} for a discussion of their relation to local thermodynamic properties of fluid phase.

\section{The non-equilibrium coset construction}\label{non-eq coset section}

Before we are able to state the procedure for constructing non-equilibrium effective actions with the method of cosets, we must first understand the IR field content of condensed matter systems at finite temperature. It turns out that Goldstone and Goldstone-like excitations at finite temperature are a bit different than at zero temperature. After the field content is established, we will outline the procedure for  using cosets to construct  non-equilibrium effective actions, which we term the {\it non-equilibrium coset construction}. 

\subsection{Goldstones at finite temperature}

At finite temperature, SSB occurs when a symmetry generator fails to commute with the equilibrium density matrix. Goldstone's theorem tells us that for every spontaneously broken symmetry, there is a corresponding Goldstone mode (unless inverse Higgs constraints are imposed). However, at finite temperature, there are other excitations that survive over long distances and extended time scales, which  resemble Goldstones. Our claim is that there are in fact Goldstone-like fields for {\it every} symmetry regardless of whether it is broken or unbroken (unless inverse Higgs-type constraints are imposed). This claim is best understood from a semi-classical perspective. Semi-classically, the density matrix is a purely pragmatic tool that is used to account for our classical ignorance of the true micro-state of the system. Essentially every (classical) micro-state in a thermal statistical ensemble corresponds to a highly chaotic classical field configuration, which will in general spontaneously break {\it every} symmetry. As a result, we expect that the non-equilibrium effective action should consist of Goldstones as if every symmetry of the theory were spontaneously broken. We will refer to the Goldstones corresponding to spontaneously broken generators as {\it broken Goldstones} and those corresponding to unbroken generators as {\it unbroken Goldstones}.

 It turns out that there is an important distinction between broken and unbroken Goldstones. In particular, in every known case, unbroken Goldstones possess infinitely many gauge symmetries. For specific examples, consult~\cite{H. Liu,H. Liu 2,H. Liu 3,Nicolis,MHD}.  While it is not fully understood from first principles where these gauge symmetries come from, Appendix \ref{Symmetry origin} gives a pragmatic explanation for why they should exist.
We will take as a well-motivated {\it assumption} that these gauge symmetries act as follows. Let $T_A$ be the unbroken generators and let $\epsilon^A_s(\phi)$ be the unbroken Goldstones.  Here, $s=1,2$ indicates on which leg of the CPT the fields live. Then we have the gauge redundancies 
\begin{equation}\label{partial gauge 0}\begin{split} 
e^{i \epsilon_s^A(\phi) T_A}&\to e^{i \epsilon_s^A(\phi) T_A} e^{i \lambda^A(\phi^I) T_A},
\end{split}\end{equation} 
for arbitrary spatial functions $\lambda^A(\phi^I)$. We use the convention that $M,N=0,1,2,3$ and $I,J=1,2,3$. 
Additionally, the effective action will be invariant under a certain subgroup of diffeomorphisms  that act on the fluid worldvolume  coordinates $\phi^M$ by~\cite{H. Liu} 
\begin{equation}\label{partial gauge 1}\begin{split} 
\phi^M & \to\phi^M+\xi^M(\phi^I),
\end{split}\end{equation} 
which we interpret as additional gauge symmetries. 
Since it is unknown how exactly these symmetries arise, it is conceivable that they may not hold in all situations. 
However, we will see in the following sections that~(\ref{partial gauge 0}) and~(\ref{partial gauge 1}) are in fact the correct symmetries for a wide range of physical systems.

\subsection{The method of cosets}

As discussed in the previous subsection, at finite temperature there is a Goldstone mode corresponding to each symmetry of the theory (unless inverse Higgs-type constraints are imposed). As a result, in the coset construction we must parameterize the full symmetry group of the theory with Goldstone modes.\footnote{Since we are parameterizing the full symmetry group as opposed to merely the non-linearly realized coset,  it is not technically correct to call our construction a {\it coset} construction. Nevertheless we will use the term `coset construction' for the sake of  linguistic continuity. } Since unbroken Goldstones enjoy a gauge symmetry whereas broken Goldstones do not, we must distinguish between the two types of Goldstones. Additionally, we assume that there exist some sorts of unbroken translation generators~\cite{Zoology}. The {\it global} symmetry generators of the theory are as follows:\footnote{When using the coset construction for theories with gauge symmetries, there are two main approaches. In \cite{coset,Joyce}, gauge transformations are treated as if they are physical symmetries and generators for each gauge transformation appear in the coset. In \cite{Wheel}, gauge symmetries are treated as redundancies and only global symmetry generators appear in the coset. Both methods yield correct results, but we will use the second approach as it is far simpler.}
\begin{equation}\begin{split}\label{the symmetry generators} \bar P_\mu & =\text{unbroken translations},
\\ T_A & = \text{other unbroken generators},
\\ \tau_\alpha & =\text{broken generators}. 
 \end{split}\end{equation}
 Finally, let $\cG$ be the full symmetry group of the theory (including spacetime symmetries), let $\cH\subset \cG$ be the unbroken subgroup, and let $\cH_0\subset \cH$ be the subgroup generated by $T_A$.
 
We will construct our theory on the fluid worldvolume with coordinates $\phi^M$ for $M=0,1,2,3$. Parameterize an arbitrary group element by 
\begin{equation}\label{coset for non-eq} g_s(\phi) = e^{i X^\m_s(\phi)\bar P_\mu} e^{i \pi^\alpha_s(\phi)\tau_\alpha} e^{i \epsilon_s^A(\phi) T_A}, \end{equation}
where $s=1,2$ indicates on which leg of the SK contour the fields live. Each $g_s$ for $s=1,2$ transforms under the same global symmetry action.  Notice that the spacetime coordinates $x^\m$ that appear in the ordinary coset construction (\ref{ordinary coset}) have been promoted to dynamical fields $X^\m_s(\phi)$ in the non-equilibrium coset construction. These fields serve to embed the fluid worldvolume into the physical spacetime. 

 We are interested in building-blocks for the non-equilibrium effective action that can be constructed from the Maurer-Cartan one-form and that transform covariantly under the global symmetry group $\cG$ as well as under the gauge transformations 
\begin{equation}\label{partial gauge}\begin{split} g_s(\phi) & \to g_s(\phi^M +\xi^M(\phi^I)),
\\ g_s(\phi)&\to g_s(\phi) e^{i \lambda^A(\phi^I) T_A},
\end{split}\end{equation} 
for arbitrary spatial functions $\xi^M(\phi^I)$ and $\lambda(\phi^I)$.\footnote{Once again, we use $M,N=0,1,2,3$ and $I,J=1,2,3$.} To this end, we will treat the coefficients of unbroken generators of $\cH_0$ in the Maurer-Cartan form as gauge-fields that have the partial gauge-invariance defined by the second line of~(\ref{partial gauge}). The  Maurer-Cartan one-form is
\begin{equation}\label{MC form} g_s^{-1} \partial_M g_s =  iE_{sM}^\mu \cur{ \bar P_\mu + \nabla_{\m} \pi_s^\alpha \tau_\alpha} +i \cB^A_{sM} T_A, \end{equation}
where $E_{sM}^\mu$ are the vierbeins, $\nabla_{\m}\pi^\alpha_s$ are the covariant derivatives of the broken Goldstones, and certain components of $\cB_{sM}^A$ behave like gauge connections.\footnote{ Notice that terms from $\log (g_1^{-1}\cdot g_2)$ are invariant under the global symmetry group $\cG$ and covariant under the gauge transformations, yet they are not used as building-blocks of the effective action. Our claim is that only building-blocks that arise from the Maurer-Cartan one-form are permissible, the demonstration of which is provided in Appendix \ref{Stückelberg}.} 
These objects transform under (\ref{partial gauge}) as follows. Under the $T_A$-gauge transformations, the vierbeins, the covariant derivatives, and $\cB_{s0}^A$ transform linearly, whereas $\cB_{sI}^A$ transform as 
\begin{equation}\begin{split} \cB_{sI}^AT_A \to \cB^A_{sI} h_0^{-1}\cdot T_A\cdot h_0 -i h_0^{-1}\cdot\partial_I h_0,\end{split}\end{equation}
where $h_0(\phi^I)\equiv e^{i \lambda^A(\phi^I) T_A}$; that is, $\cB_{sI}^A$ transform as connections. Under the fluid diffeomorphism transformations, $E_{s0}^\m$, $\nabla_{\m}\pi_s^\alpha$, and $\cB_{s0}^A$ all transform as scalars, whereas the spatial component of the vierbeins and the connections transform as
\begin{equation}\begin{split}
E_{sI}^\m \to \frac{\partial \phi^M}{\partial \phi'^I} E_{sM}^\m,
\\ \cB_{sI}^A \to \frac{\partial \phi^M}{\partial \phi'^I} \cB_{sM}^A,
\end{split}\end{equation}
where $\phi'^M\equiv \phi^M+\xi^M(\phi^I)$. 

The building-blocks that transform covariantly under both the global symmetries and the gauge symmetries~(\ref{partial gauge}) are as follows: First, there are the building-blocks from the usual coset construction, namely $\nabla_{\m } \pi^\alpha_s$, which transform covariantly under~(\ref{partial gauge}),  and  to take higher-order covariant derivatives we can use 
\begin{equation}\label{cov. deriv.ative} \frac{\partial}{\partial \phi^0},~~~~~~~~~~\nabla^\cH _{I} = {\partial_I + i \cB_{rI}^A T_A}, \end{equation}
where $\cB_{rI}^A =\frac{1}{2}\cur{\cB_{1I}^A+\cB_{2I}^A} $ and $I=1,2,3$. 
 To contract coordinate indices, we use the metrics
\begin{equation}G_{sMN} =  E^\m_{sM} \eta_{\m\n} E^\n_{sN}. \end{equation} 
Second, there are new building-blocks that involve the unbroken Goldstone degrees of freedom, 
namely $E_{s0}^\mu$ and $\cB_{s0}^A$, which transform covariantly. Finally, we have terms that involve combinations of 1-and 2-fields. Notice that $E_{1M}^\m (E_2^{-1})^M_\n$ and $\cB_{aM}^A\equiv \cB_{1M}^A-\cB_{2M}^A$ transform covariantly and we can contract coordinate indices with $E^\m_{1M}\eta_{\m\n}E^\n_{2N}$.

After computing these building-blocks, inverse Higgs-type constraints can be imposed to remove extraneous Goldstone modes. The basic idea behind inverse Higgs constraints is that we may set to zero any objects that transform covariantly under the symmetries and gauge symmetries as long as the resulting equations can be algebraically solved for one set of Goldstones in favor of derivatives of another set of Goldstones. We will see in subsequent examples that there are three kinds of inverse  Higgs-type constraints. 

There are the usual inverse Higgs constraints that exist for zero-temperature systems. Suppose that the commutators between an unbroken translation generator $\bar P$ and a broken generator $\tau '$ contains another unbroken generator $\tau$, that is $[\bar P, \tau '] \supset \tau$. Further, suppose that $\tau$ and $\tau '$ do not belong to the same irreducible multiplet under $\cH_0$. Then, it is consistent with symmetry transformations to set the covariant derivative of the $\tau$-Goldstone in the direction of $\bar P$ to zero. This gives a constraint that relates the $\tau '$-Goldstone to derivatives of the $\tau$-Goldstone, allowing the removal of the $\tau'$-Goldstone. 

Notice that the ordinary inverse Higgs constraints only allow for the removal of certain broken Goldstones by relating them  to derivatives of other broken Goldstones. One might wonder if there are other possibilities. Perhaps we could remove broken Goldstones by relating them to derivatives of unbroken Goldstones; or perhaps we could specify some constraints that allow us to remove unbroken Goldstones. It turns out that both of these are possible; the rules are as follows:\footnote{To get a better understanding of why these new inverse Higgs constraints can be imposed, see Appendix~\ref{explicit IH computations}. In this appendix, to keep things as concrete as possible, we perform an in-depth analysis of inverse Higgs-type constraints for fluids without conserved charge.}
\begin{itemize}
\item {\it Thermal inverse Higgs:} Suppose that at finite temperature, the commutator between a broken generator $\tau$ and the unbroken time-translation generator $\bar P_0$ contains an unbroken spacetime translation generator $\bar P$, that is $[\tau,\bar P_0]\supset \bar P$. Then we may set to zero the component of $E_0^\m$ in the direction of $\bar P$. This gives an equation that can be algebraically solved to yield an expression for the $\tau$-Goldstone in terms of derivatives of the $\bar P$-Goldstone. This allows the removal of the $\tau$-Goldstone. 
\item {\it Unbroken inverse Higgs:} Suppose that at finite temperature, the commutator between an unbroken generator $T$ and an unbroken spacetime translation generator $\bar P'$ contains another unbroken spacetime translation generator $\bar P $, that is $[T,\bar P']\supset \bar P$. Consider the matrix ${A^\m}_\n\equiv  (E_1)_M^\m (E_2^{-1})^M_\n$, where $M=0,1,2,3$ are coordinate indices and $\m,\n=0,1,2,3$ are Lorentz indices. Then we may set to zero the components of $A_{\m\n}$ in the directions of $\bar P$ and $\bar P' $. Suppose that under the dynamical KMS symmetry transformation, $A_{\m\n}\to \tilde A_{\m\n}$. Then, we may also set to zero the components of $\tilde A_{\m\n}$ in the directions of $\bar P$ and $\bar P' $. These conditions give constraints that relate the $T$-Goldstone to derivatives of the $\bar P$-Goldstone, allowing the removal of the $T$-Goldstone.\footnote{It turns out that the unbroken inverse Higgs constraints do not necessarily remove the unbroken Goldstones from the {\it covariant} building-blocks. However they do remove the unbroken Goldstones entirely from the {\it invariant} building-blocks and hence from the effective action; see Appendices \ref{c2}-\ref{c3}.}
\end{itemize}

One might wonder if the aforementioned inverse Higgs constraints are the only possible such constraints. From a very general stand-point, an inverse Higgs constraint can be imposed any time there is a covariant object that when set to zero, results in an equation which can be algebraically solved for one set of Goldstones in terms of another. With this level of generality, it is hard to say if we have been exhaustive. However, most inverse Higgs constraints are related to commutator relations of the form $[A,\bar P]\supset B$, for unbroken translation generator $\bar P$ and some generators $A$ and $B$. Because the set of unbroken generators forms a closed algebra, it is impossible for $B$ to be broken yet $A$ unbroken. Every other combination of broken and  unbroken, however, is permitted. If both $A$ and $B$ are broken, then we have ordinary inverse Higgs constraints;  if $A$ is broken and $B$ unbroken, we have thermal inverse Higgs constraints; and if both $A$ and $B$ are unbroken, we have unbroken inverse Higgs constraints. Thus, our list of inverse Higgs constraints is exhaustive if we assume that all such constraints must be related to commutator relations.\footnote{Strictly speaking, the thermal and unbroken inverse-Higgs constraints make the assumption that $B$ is an unbroken translation generator. It is possible to conceive of more general possibilities for $B$, but such inverse Higgs constraints will not be relevant for any constructions in this paper.}  

Once inverse Higgs constraints have been imposed and an effective action has been constructed, it  is then necessary to impose the dynamical KMS symmetry. Often, at leading order in the derivative expansion (which is the only order to which we will be working in the subsequent examples) the only effect of the dynamical KMS symmetry will be to mandate that the effective action factorize as
\begin{equation}\label{factorized eft} I_\text{EFT}[\chi_r,\chi_a] = S_\text{EFT}[\chi_1]-S_\text{EFT}[\chi_2]+\cO(\chi_ a^3), \end{equation}
where $\chi_r\equiv \frac{1}{2}\cur{\chi_1+\chi_2}$,  $\chi_a\equiv {\chi_1-\chi_2}$, and $\cO(\chi_ a^3)$ counts as higher order in the derivative expansion. The only exceptions to this rule that appear in this paper are the EFTs for nematic and smectic C phases of liquid crystals. 
When this rule holds, it allows us to work with just one copy of the fields. Thus, at leading order, the building-blocks  $\nabla_\m\pi^\alpha$  describe the broken Goldstones and $E^\mu_0$ and $\cB^A_0$  describe the unbroken Goldstones. For the sake of simplicity, in the subsequent sections, we will construct EFTs for various states of matter at leading order in the derivative expansion. As a result, we will construct ordinary actions with just one copy of the fields whenever possible.
 
 It should be noted that the primary reason for defining effective actions on the SK contour is to account for dissipation. When an action factorizes according to~(\ref{factorized eft}), the leading-order physics is fully captured by an ordinary action and is thus non-dissipative. Therefore, the majority of the effective actions constructed in this paper will not account for dissipative effects. However, our coset construction gives a clear, almost mechanical prescription for constructing higher-order terms, which if included in the EFT, will account for dissipation. We leave this important work of computing higher-order terms for future study. 

\section{Fluids and superfluids at finite temperature}

To demonstrate that our non-equilibrium coset construction presented in the previous section gives correct results, we will reproduce the known effective actions for fluids and superfluids at finite temperature.  Along the way, we will find that finite-temperature framids---systems for which only boosts are spontaneously broken~\cite{Zoology}---can be thought of as fluids. In particular, at finite temperature, the boost Goldstones automatically become gapped and can therefore be integrated out, resulting in the ordinary fluid EFT. Finally, the equations of motion for fluids and superfluids from the EFT perspective have been studied in great detail in~\cite{H. Liu,Nicolis, Finite T superfluid}. Thus, we will not study them here. For all other states of matter, however, we will include brief discussions of the resulting equations of motion. 

\subsection{Fluids}\label{Fluids}

Consider a fluid without a conserved charge; the symmetry group is just the Poincaré group, whose algebra is given by~(\ref{Poincare}). The unbroken generators are $P_\mu$ for translations and $J_{i}\equiv \frac{1}{2}\epsilon^{ijk} J_{jk}$ for spatial rotations; the broken generators are $K_i\equiv J_{0i}$ for boosts. Therefore, the most general group element is
\begin{equation} g(\phi) = e^{iX^\mu(\phi)P_\m} e^{i\eta^i(\phi) K_i} e^{{i} \theta^{i}(\phi) J_{i}}. \end{equation}
The resulting Maurer-Cartan one-form is 
\begin{equation} g^{-1}\partial_M g = i E^\mu_M \cur{ P_\mu +\nabla_\m \eta^i K_i}+i \Omega_M^{i} J_{i},  \end{equation}
where the vierbein, covariant derivative, and spin-connection are given by
\begin{equation}\begin{split}\label{vierbein spin} E^\mu_M &= \partial_M X^\nu {[\Lambda R]_\nu}^\m,
\\ \nabla_\m \eta^i & = (E^{-1})^M_\mu [ \Lambda^{-1}\partial_M \Lambda  ]^{0j} R^{ji},
\\ \Omega_M^{i} &=\frac{1}{2}\epsilon^{ijk} [R^{-1} \Lambda^{-1}\partial_M (\Lambda R) ]^{jk},
\end{split} \end{equation}
such that $R^{ij}=[e^{{i} \theta^{i}(\phi)J_{i}}{]^{ij}}$ and ${\Lambda^\m}_\n=[e^{{i} \eta^{i}(\phi)K_{i}}{]^\m}_\n$. 
Because spatial rotations are unbroken, the action must be invariant under the transformations
\begin{equation}\begin{split}\label{fluid gauge}
g(\phi)&\to g(\phi^M+\xi^M(\phi^I)),
\\ g(\phi) &\to g(\phi) \cdot e^{{i} \lambda^{i}(\phi^I)J_{i}},
\end{split}\end{equation}
for arbitrary spatial functions $\xi^M(\phi^I)$ and $ \lambda^{i}(\phi^I) $. 
Notice that $[P_i,K_j] = i \delta_{ij} P_0$, meaning that we may impose the thermal inverse Higgs constraints, allowing the removal of $\eta^i$. In particular $E_0^i$ transforms covariantly under both the global symmetry group $\cG$ as well as the gauge and diffeomorphism symmetries~(\ref{partial gauge}). Therefore we may fix
\begin{equation}\label{thermal IH ordinary fluids}0 = E_0^i = \frac{ \partial X^\m}{\partial \phi^0} {\Lambda_\m}^j R^{ji}. \end{equation}
Since $R^{ij}$ is invertible, this is equivalent to $\partial_0 X^\m {\Lambda_\m}^i=0$, from which we find that 
\begin{equation}\label{thermal IH ordinary fluids...} \frac{\eta^i}{\eta} \tanh \eta = -\frac{\partial_0 X^i}{\partial_0 X^t},  \end{equation}
where $\eta \equiv \sqrt{\eta^i\eta^i}$. For a more detailed calculation, see Appendix \ref{c1}.

Now we can impose unbroken inverse Higgs constraints to remove the rotation Goldstones.\footnote{At this level in the derivative expansion, it is not actually necessary to solve the unbroken inverse Higgs constraints, but we will do so just to demonstrate that it can be done. When imposing the unbroken inverse Higgs constraints in later sections, such calculations will be omitted. } In order to do this, we must recall that the field-content is doubled. Notice that $[P_i,J_j]=i\epsilon_{ijk} P_k$, so we may set 
\begin{equation}\label{unbroken IH ordinary fluids} (E_1)_{i M} (E_2^{-1})^M_j-(E_1)_{jM} (E_2^{-1})^M_i=0,\end{equation}
 which gives 
 \begin{equation}\label{unbroken IH ordinary fluids...} R_1\cdot R_2^{-1} = \sqrt{\cM^T\cdot\cM}\cdot \cM^{-1},~~~~~~~~~~ \cM^{ij}\equiv (\Lambda_2^{-1}{)}^{i\n} \frac{\partial X_1^\m}{\partial X_2^\n} (\Lambda_1{)_\m}^j.\end{equation} See Appendix \ref{c2} for more detailed calculations. To get an intuitive sense of what is happening, let us expand the above equation to linear order in the fields. Transforming from the 1,2-basis to the $r,a$-basis~(\ref{r a variables}), we have
\begin{equation} \vec\theta_a = \p\times\vec X_a,  \end{equation} 
where $\p\times$ represents the ordinary curl and {\it not} a covariant derivative. Then, performing the (classical) dynamical KMS transformation on both sides gives
\begin{equation}\label{linearized unbroken inverse Higgs}\frac{\partial\vec\theta_r}{\partial\phi^0} = \p\times \frac{\partial \vec X_r}{\partial\phi^0} . \end{equation} 
But notice that the linearized version of~(\ref{fluid gauge}) implies that our action must be invariant under $\vec\theta_r (\phi) \to \vec \theta_r(\phi)+\vec\lambda(\phi^I)$,
 meaning that at linear order, $\vec\theta_r$ can only appear in the effective action in the form $\partial \vec\theta_r/\partial \phi^0$. As a result,~(\ref{linearized unbroken inverse Higgs}) is sufficient to remove $ \vec \theta_r$ as an independent degree of freedom. Thus, $\vec\theta_r$ and $\vec\theta_a$ have been successfully removed from the effective action, so no rotational Goldstones remain. To see how $\vec \theta_r$ can be removed at the non-linearized level, see Appendix \ref{c3}. 

Thus the only covariant building-block at leading order in the derivative expansion is $E^t_0 = \sqrt{- E^\m_0\eta_{\m\n} E^\n_0}$.\footnote{To avoid ambiguity, instead of writing $E^0_0$, we replace the Lorentz index with a $t$. This way it is clear that $E^t_0$ has one raised Lorentz index and one lowered coordinate index. Whenever we feel that there may be an ambiguity, we will use $\mu=t$ instead of $\m=0$ to indicate time-like components of Lorentz vectors and tensors.} Defining the metric by\footnote{The metric tensor defined in~(\ref{bimetric}) agrees exactly with the pull-pack fluid metrics of~\cite{H. Liu}.}
\begin{equation}\label{bimetric}G_{MN} \equiv E^\mu_M\eta_{\m\n} E^\n_N = \frac{\partial X^\m}{\partial\phi^M}\eta_{\m\n} \frac{\partial X^\n}{\partial\phi^N},\end{equation}
we find that the leading-order action is a generic function of $G_{00}=-(E_0^t)^2$ and is given by
\begin{equation}\label{zeroth order fluid action} S_\text{EFT} = \int d^4 \phi \sqrt{-G} ~P(G_{00}).  \end{equation}

Notice that the above system has the same symmetry-breaking pattern as a framid except for the fact that it exists at finite temperature~\cite{Zoology}. Thus, we can interpret a finite-temperature framid as a fluid. With this interpretation, the meaning of the thermal inverse Higgs constraint is as follows. If we did not set $E_0^i=0$, then we could use it as a covariant building-block. But $E_0^i\supset \eta^i +\cdots$, meaning that terms involving $\eta^i$ with no derivatives must exist in the action. Therefore $\eta^i$ has an energy gap.\footnote{At finite temperature, the notion of an energy gap is ill-defined. Really, what we mean is that the propagator will die off in space and time exponentially fast. But since the leading-order hydrodynamic action takes the form of an ordinary action with an energy gap, we use the term `energy gap.' } Thus at finite temperature, framid Goldstones necessarily develop a gap and can therefore be integrated out. It should be noted, however, that at sufficiently low temperatures, the framid Goldstones' energy gap may be quite small. In this case it would only be appropriate to integrate them out if we are interested exclusively  in the very deep IR behavior of the system. 

Now suppose that the fluid carries a conserved  $U(1)$  charge $Q$ and exists at finite chemical potential. Then the unbroken translations are $\bar P_0=P_0+\m Q$ and $\bar P_i=P_i$.\footnote{Notice that both $P_0$ and $Q$ are unbroken, so defining $\bar P_0=P_0+\m Q$ reflects a choice since any linear combination of $P_0$ and $Q$ is unbroken. But it is the most natural choice given that this definition of $\bar P_0$ allows us to express the equilibrium density matrix in the usual form $\rho=e^{-\beta_0 \bar P^0}/\tr e^{-\beta_0 \bar P^0} $. } Only boosts are spontaneously broken and the most general group element is 
\begin{equation} g(\phi) = e^{iX^\mu(\phi)\bar P_\m} e^{i\pi(\phi) Q} e^{{i} \eta^{i}(\phi) K_{i}} e^{{i} \theta^{i}(\phi) J_{i}}. \end{equation}
The resulting Maurer-Cartan one-form is
\begin{equation} g^{-1}\partial_M g = i E^\mu_M \cur{ \bar P_\mu +\nabla_\m \eta^i K_i} +i \cB_M Q+{i} \Omega_M^{i} J_{i},  \end{equation}
where the vierbein, covariant derivative, and spin-connection are given by~(\ref{vierbein spin}) and the $U(1)$ gauge field is given by
\begin{equation}\label{gauge field expression} \cB_M = \partial_M \psi-\m E^t_M ,\end{equation} 
where $\psi(\phi)\equiv \m X^t(\phi)+\pi(\phi)$.
We are interested in constructing an action that is invariant under~(\ref{fluid gauge}) as well as the chemical shift gauge symmetry 
\begin{equation}\begin{split}\label{fluid gauge u(1)}
g(\phi) &\to g(\phi) \cdot e^{\lambda_Q(\phi^I)Q},
\end{split}\end{equation}
for arbitrary spatial function $\lambda_Q(\phi^I)$. 
As before, we may remove $\eta^i$ by imposing  thermal inverse Higgs constraints~(\ref{thermal IH ordinary fluids}), and we may remove $\theta^i$ by imposing unbroken inverse Higgs constraints~(\ref{unbroken IH ordinary fluids}). 

Notice  that the $U(1)$ gauge field $\cB_0$ transforms as a scalar. Since $E^t_0$ also transforms as a scalar, it is convenient to  define the invariant building-block  $B_0\equiv \cB_0+\m E^t_0=\partial_0\psi$. The effective action at lowest order in derivatives is therefore 
\begin{equation}\label{zeroth order fluid action with charge} S_\text{EFT} = \int d^4 \phi \sqrt{-G}  ~P(B_0,G_{00}). \end{equation}
Notice that~(\ref{zeroth order fluid action}) and~(\ref{zeroth order fluid action with charge}) are precisely the leading-order actions presented in~\cite{H. Liu,H. Liu 2,H. Liu 3}. They appear to differ from the actions presented in~\cite{Nicolis}, but it can be shown that they are equivalent; see Appendix~\ref{V diff}.

\subsection{Superfluids}

Consider a superfluid at finite temperature. In addition to possessing Poincaré symmetry~(\ref{Poincare}), such a system must also have a conserved $U(1)$ charge $Q$ such that 
$ \bar P_0  = P_0 + \mu Q$, $ \bar P_i  = P_i$,   and
$J_i  = \frac{1}{2} \epsilon_{ijk} J_{jk}$
are the unbroken generators and 
$  Q$ and $ K_i  = J_{0i}$
are the broken generators~\cite{Zoology,Finite T superfluid}. The most general group element is 
\begin{equation} g(\phi) = e^{i X^\m(\phi)\bar P_\mu} e^{i\pi(\phi) Q} e^{i\eta^i(\phi) K_i} e^{i\theta^i(\phi) J_i}.\end{equation}
And the resulting Maurer-Cartan one-form is 
\begin{equation}g^{-1}\partial_M g = i E^\m_M\cur{\bar P_\mu + \nabla_\m \pi Q+\nabla_\m \eta^i K_i} + i \Omega^i_M J_i, \end{equation} 
where 
\begin{equation}\begin{split} E^\m_M&= \partial_M X^\n {[\Lambda R]_\n}^\m,
\\ \nabla_\mu \pi & = (E^{-1})^M_\m \partial_M\psi - \mu \delta_\m^0,
\\ \nabla_\m \eta^i & = (E^{-1})^M_\mu [ \Lambda^{-1}\partial_M \Lambda  ]^{0j} R^{ji},
\\ \Omega^i_M & = \frac{1}{2} \epsilon^{ijk}[R^{-1}\Lambda^{-1}\partial_M (\Lambda R)]^{jk},
 \end{split} \end{equation} 
such that $\psi (\phi)\equiv \m X^0(\phi)+\pi(\phi)$, the Lorentz boost matrix is ${\Lambda^\m}_\n(\phi) = [e^{i\eta^i(\phi) K_i}{]^\m}_\n$, and the rotation matrix is $R^{ij}(\phi) = [e^{i\theta^k(\phi) J_k}]^{ij}$. To remove the boost Goldstones, impose inverse Higgs constraints $0=\nabla_i\pi={\Lambda^\rho}_i (e^{-1})_\rho^M \partial_M\psi$, where $e_M^\m \equiv \partial_M X^\m$. These constraints give a relation between the boost Goldstones and the $U(1)$ Goldstones
\begin{equation}\label{IH relation}  \frac{\eta^i}{\eta} \tanh {\eta}= -\frac{(e^{-1})^M_i \partial_M\psi}{(e^{-1})^M_0 \partial_M\psi}.\end{equation}
Using this relation, we have that $\nabla_t\pi = \sqrt{-G^{MN}\partial_M\psi\partial_N\psi}-\m$, giving our first symmetry-invariant building-block
\begin{equation}Y\equiv  \sqrt{-G^{MN}\partial_M\psi\partial_N\psi}. \end{equation}
Equation~(\ref{IH relation}) tells us that $[\Lambda R{]^t}_\mu = \frac{1}{Y} (e^{-1})^M_\m \partial_M\psi$, from which we find that $E^\mu_0\nabla_\mu \pi =[1-\frac{\mu}{Y}]\partial_0 \psi$, yielding our second symmetry-invariant building-block
\begin{equation} B_0\equiv \frac{\partial \psi}{\partial \phi^0}. \end{equation}
And just as in the fluid case, we have
\begin{equation}G_{00} = E^\mu_0 \eta_{\m\n} E^\n_0 \end{equation}
as our third symmetry-invariant building-block. 
Finally, following the same calculation as in the previous section, we can impose unbroken inverse Higgs constraints to remove the rotation Goldstones. Thus, the effective action describing a superfluid at finite temperature is, to lowest order in derivatives, 
\begin{equation}\label{coset finite T superfluid} S_\text{EFT} = \int d^4 \phi \sqrt{-G} ~P(Y,B_0,G_{00}).  \end{equation} 
This effective action looks somewhat different from the one presented in~\cite{Finite T superfluid}, but can be shown to be equivalent; see Appendix \ref{V diff}. Further, in the zero-temperature limit, the effective action loses its dependence on $X^\m(\phi)$, meaning that it only depends on the physical-spacetime version of $Y$, namely~(\ref{zero T y}), reproducing the standard superfluid EFT~(\ref{zero T superfluid EFT}).

\section{Solids and supersolids at finite temperature} 

Now that we have established that the non-equilibrium coset construction can reproduce known results, we turn our attention to the construction of novel EFTs. The simplest physical systems for which the non-equilibrium effective actions are as of yet unknown are solids and supersolids. In this section, we will construct the leading-order non-equilibrium EFTs for these states of matter. 

\subsection{Solids}\label{Solids}

Consider a chargeless crystalline solid at finite temperature. In addition to possessing Poincaré symmetry~(\ref{Poincare}), such a system must also have internal translation symmetry generators $Q_i$ for $i=1,2,3$ that commute with each other. We then take
$\bar P_0 = P_0$ and $\bar P_i =P_i+Q_i$
to be the unbroken generators such that 
$Q_i$, $K_i=J_{0i}$, and $J_i = \frac{1}{2} \epsilon_{ijk}J_{jk}$ 
are the broken generators~\cite{Zoology}. The most general group element is
\begin{equation} g(\phi)=e^{i X^\mu(\phi)\bar P_\mu} e^{i\pi^i(\phi)Q_i} e^{i\eta^i(\phi)K_i}e^{i\theta^i(\phi)J_i}. \end{equation}
And the resulting Maurer-Cartan one-form is 
\begin{equation}\label{Solid MC form}g^{-1}\partial_M g = i E^\m_M\cur{\bar P_\mu + \nabla_\m \pi^i Q_i+\nabla_\m \eta^i K_i+\nabla_\mu\theta^i J_i} , \end{equation} 
where 
\begin{equation}\begin{split} E^\m_M&= \partial_M X^\n {[\Lambda R]_\n}^\m,
\\ \nabla_\mu \pi^i & = (E^{-1})^M_\m \partial_M\psi^i - \delta_\m^i,
\\ \nabla_\m \eta^i & = (E^{-1})^M_\mu [ \Lambda^{-1}\partial_M \Lambda  ]^{0j} R^{ji},
\\ \nabla_\m \theta^i & = (E^{-1})^M_\m \epsilon^{ijk}[R^{-1}\Lambda^{-1}\partial_M ( \Lambda R)]^{jk},
 \end{split} \end{equation} 
such that $\psi^i (\phi)\equiv X^i(\phi)+\pi^i(\phi)$, the Lorentz boost matrix is ${\Lambda^\m}_\n(\phi) \equiv [e^{i\eta^i(\phi) K_i}{]^\m}_\n$, and the rotation matrix is $R^{ij}(\phi) \equiv  [e^{i\theta^k(\phi) J_k}]^{ij}$. To remove the boost and rotation Goldstones, impose inverse Higgs constraints $\nabla_t\pi^i =0$ and $\p\times \vec\pi=0$, respectively.\footnote{Here, $\p$ should be thought of as the spatial components of the covariant derivative $\nabla_\mu$. Thus $\p\times \vec\pi$ is {\it not} the curl of $\vec \pi$. } Let $e^\m_M\equiv \partial_M X^\m$. Then by imposing the first inverse Higgs constraint, we find that 
\begin{equation}\label{IH relation solid 1}  \frac{\eta^j}{\eta} \tanh {\eta}= -(e^{-1})_t^M \partial_M \psi^j (a^{-1})^{ij}   \end{equation}
where $a^{ij}\equiv (e^{-1})_i^M\partial_M \psi^j$. Imposing the second inverse Higgs constraint tells us that 
 $\nabla^{(i} \pi^{j)} = (Y^{1/2})^{ij} -\delta^{ij}$, where 
\begin{equation} Y^{ij}\equiv G^{MN} \partial_M\psi^i\partial_N\psi^j,  \end{equation}  
such that  $G^{MN}$ is the inverse of the metric~(\ref{bimetric}).
We therefore have our first set of invariant building-blocks. Notice that because there is no notion of unbroken rotational symmetry, $Y^{ij}$ is truly symmetry-invariant and not merely covariant. We have the additional invariant objects $E_0^\mu \nabla_\mu\pi^i = (\delta^{ij} -(Y^{-1})^{ij}) \partial_0 \psi^j$, from which we extract our next set of invariant building-blocks
\begin{equation} Z^i \equiv \frac{\partial \psi^i}{\partial\phi^0}.  \end{equation}
Lastly, we have the usual time-like piece of the fluid metric~(\ref{bimetric}), $G_{00}$, as our final building-block. Thus the effective action describing a crystalline solid at finite temperature is, to lowest order in derivatives,
\begin{equation}\label{solid EFT no charge} S_\text{EFT} = \int d^4 \phi\sqrt{-G}~P(Y^{ij},Z^i,G_{00}).  \end{equation}

Often, solids possess an additional unbroken $U(1)$ symmetry corresponding to particle-number conservation. Let $Q$ be the corresponding generator. Then, the unbroken translations are $\bar P_0=P_0+\m Q$  and $\bar P_i=P_i+Q_i$. The most general group element is
\begin{equation} g(\phi)=e^{i X^\mu(\phi)\bar P_\mu} e^{i\pi^i(\phi)Q_i} e^{i\eta^i (\phi)K_i}e^{i\theta^i(\phi)J_i} e^{i\pi (\phi)Q}. \end{equation}
The corresponding Maurer-Cartan one-form is just~(\ref{Solid MC form}) with the addition of a term involving the $U(1)$ gauge-field $\cB_M = \partial_M\psi-\m E^t_M$ for $\psi(\phi)=X^0(\phi)+\pi(\phi)$. As when we added a conserved charge to fluids, we now have the additional building-block  $B_0 \equiv \partial_0 \psi$. Thus the effective action describing a crystalline solid with conserved $U(1)$ charge at finite temperature is, to lowest order in derivatives,
\begin{equation} \label{charged solid eft} S_\text{EFT} = \int d^4 \phi\sqrt{-G}~P(Y^{ij},Z^i,B_0,G_{00}).  \end{equation}
As a consistency check, let us take the zero-temperature limit. This amounts to removing all dependence on ${\partial X^\mu(\phi)}/{\partial\phi^0}$. We see therefore that the zero-temperature EFT only depends on  the physical-spacetime version of $Y^{ij}$, namely
\begin{equation}y^{ij}\equiv \partial_\m \psi^i\partial^\mu\psi^j,\end{equation}
 which agrees with the results of \cite{coset}.

To connect with other formulations of the finite-temperature hydrodynamics of solids~\cite{solid ref 1,solid ref 2,solid ref 3, solid ref 4,unified hydro}, we compute the equations of motion. If we vary the chargeless action with respect to $X^\mu$, we find the conservation of the stress-energy tensor $\partial_\nu T^{\mu\nu}=0$. The stress-energy tensor is given by
 \begin{equation}\label{solid eom 1} T^{\mu\nu} = \varepsilon u^\mu u^\mu + p \Delta^{\mu\nu} + r^{\mu\nu}, ~~~~~~~~~~\Delta^{\mu\nu}\equiv \eta^{\mu\nu}+u^\mu u^\nu, \end{equation}
 where 
 \begin{equation}\label{solid eom 2} p=P,~~~~~~~~~~\varepsilon = -2 G_{00} \frac{\partial P}{\partial G_{00}} - P,~~~~~~~~~~u^\mu =\frac{1}{\sqrt{-G_{00}}} \frac{\partial X^\mu}{\partial \phi^0}\end{equation}
are the thermodynamic pressure, energy density, and four-velocity, respectively and 
\begin{equation}\label{solid eom 3} r_{\mu\nu} = \frac{\partial P}{\partial Y^{ij}} \partial_\mu \psi^i \partial_\nu \psi^j \end{equation} 
is the elastic stress-tensor. So far, our theory appears to agree with standard results. However, we have the additional equations of motion that come from varying $\psi^i$,
\begin{equation}\label{solid additional eom} \partial_\mu J^{i\mu}  = 0, ~~~~~~~~~~J^{i\mu}\equiv  2\partial^\mu\psi^j \frac{\partial P}{\partial Y^{ij}} + u^\mu \frac{\partial P}{\partial Z^i} . \end{equation} 
These new equations indicate  that the solid degrees of freedom can be excited without affecting the stress-energy tensor. The physical interpretation is that our solid exhibits second sound modes, analogous to those found in finite-temperature superfluids. The ordinary sound modes of solids consist of transverse and longitudinal vibrational modes of the lattice. However, if a finite-temperature crystalline solid has a sufficiently pristine lattice structure and Umklapp scattering events can be ignored, then there will be a bath of thermalized solid phonons that behave as a gas through which an additional fluid-like longitudinal phonon can propagate~\cite{solid second sound ref}. The additional equations of motion~(\ref{solid additional eom}) account for the fact that the solid degrees of freedom can move independently of the phonon gas. Further, in the standard formulation, the thermoelastic variables depend on the temperature and a symmetric tensor, which we can identify with $(-G_{00})^{-1/2}/\beta_0$ and $Y^{ij}$ respectively, where $\beta_0$ is the equilibrium inverse temperature. However, in our formalism, there is an additional quantity $Z^i$. Just like the additional equations of motion, this quantity owes its existence to the presence of second sound. In particular, it is non-zero when the phonon gas flows with respect to the solid lattice. We leave it as further work to determine how to remove second sound from solid EFTs.

\subsection{Supersolids}

Consider an anisotropic supersolid at finite temperature. In addition to possessing Poincaré symmetry~(\ref{Poincare}), such a system must also have internal translation symmetry generators $Q_\mu$ for $\mu=0,1,2,3$ that commute with each other such that
$\bar P_\mu =P_\mu+Q_\mu$
are the unbroken generators and 
$Q_\mu$ and $J_{\mu\nu}$ 
are the broken generators~\cite{Zoology}. The most general group element is
\begin{equation} g(\phi)=e^{i X^\mu(\phi)\bar P_\mu} e^{i\pi^\mu(\phi)Q_\mu} e^{\frac{i}{2}\theta^{\m\n}(\phi)J_{\m\n}}. \end{equation}
And the resulting Maurer-Cartan one-form is 
\begin{equation}\label{supersolid MC form}g^{-1}\partial_M g = i E^\m_M\cur{\bar P_\mu + \nabla_\m \pi^\n Q_\n+\frac{1}{2}\nabla_\m \theta^{\n\lambda} J_{\n\lambda} } , \end{equation} 
where 
\begin{equation}\begin{split} E^\m_M&= \partial_M X^\n {\Lambda_\n}^\m,
\\ \nabla_\mu \pi^\n & = (E^{-1})^M_\m \partial_M\psi^\n - \delta_\m^\n,
\\ \nabla_\mu \theta_{\n\lambda}  & = (E^{-1})^M_\m [\Lambda^{-1}\partial_M\Lambda]_{\n\lambda},
 \end{split} \end{equation} 
such that $\psi^\mu (\phi)= X^\mu(\phi)+\pi^\mu(\phi)$ and the Lorentz transformation matrix is ${\Lambda^\m}_\n(\phi) = [e^{i\eta^i(\phi) K_i}\cdot e^{i\theta^k(\phi) J_k} {]^\m}_\n$. To remove the Lorentz Goldstones, impose inverse Higgs constraints 
\begin{equation} 0=\nabla_{[\mu}\pi_{\n]}= {\Lambda^\rho}_\mu (e^{-1})_\rho^M\partial_M\psi_\nu   -(\m \leftrightarrow \n). \end{equation}
Letting $\cM_{\m\n} \equiv (e^{-1})_\m^M\partial_M \psi_\nu $, the inverse Higgs constraints merely require that $\Lambda^T \cdot \cM$ be a symmetric matrix, where factors of  $\eta_{\m\n}$ have been suppressed. Using this result, we have that $\nabla_{(\m}\pi_{\nu)}= (Y^{1/2})_{\m\n}-\eta_{\m\n} $, where
\begin{equation} Y^{\m\n} =  G^{MN} \partial_M\psi^\m\partial_N\psi^\n   \end{equation}
are our first set of symmetry-invariant building-blocks and we use the convention that 
\begin{equation} ({Y}^{1/2})_{\m\rho}\eta^{\rho\lambda}({Y}^{1/2})_{\lambda\nu} = Y_{\m\n}.\end{equation}
 Additionally, we have $E_0^\n \nabla_\n \pi^\mu = (\delta^\m_\n-(Y^{-1})^\m_\n) \partial_0\psi^\n$. Our next set of invariant building-blocks is therefore 
\begin{equation} Z^\mu \equiv \frac{\partial\psi^\mu}{\partial\phi^0}.  \end{equation} 
And finally, we have the usual $G_{00}$ building-block. Thus the effective action describing a crystalline supersolid at finite temperature is, to lowest order in derivatives,
\begin{equation} S_\text{EFT} = \int d^4 \phi\sqrt{-G}~P(Y^{\m\n},Z^\m,G_{00}).  \end{equation}
As in the solid EFT case, taking the zero-temperature limit amounts to removing all dependence on $\partial X^\mu(\phi)/\partial\phi^0$. Thus at zero temperature, the effective action can only depend on the physical-spacetime version of $Y^{\m\n}$, namely
\begin{equation}y^{\m\n}\equiv \partial_\rho\psi^\mu\partial^\rho\psi^\n, \end{equation}
 which agrees with the results of \cite{coset}. 
 
 Finally, the equations of motion are very similar to those of solids. In particular if we replace the $\psi^i$ for $i=1,2,3$ with $\psi^\mu$ for $\mu=0,1,2,3$ in equations (\ref{solid eom 1}) and (\ref{solid additional eom}), we find the finite-temperature supersolid equations of motion are $\partial_\nu T^{\mu\nu}=0$ and $\partial_\nu J^{\mu\nu} = 0$.

\section{Liquid crystals at finite temperature}\label{liquid crystals}

\begin{SCfigure}[0.57][h]
\caption{\small The figure to the left depicts the microscopic appearance of various phases of liquid crystal in order from lowest to highest temperature. The phases are (a) crystalline solid, (b) smectic liquid crystal in phase C, (c) smectic liquid crystal in phase A, (d) nematic liquid crystal, and (e) isotropic, i.e. fluid phase. Notice that (a) spontaneously breaks the most symmetries and each subsequent phase spontaneously breaks fewer and few symmetries until (e) only breaks boosts.}\label{fig 1}
\includegraphics[width=0.6\textwidth]{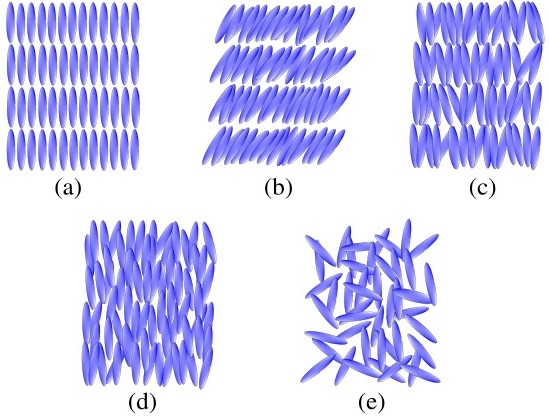}
\end{SCfigure}

We now turn our attention to the construction of EFTs for more exotic states of matter, namely those of liquid crystals. There are myriad distinct phases of liquid crystals, so for the sake of brevity (and the reader's patience) we will focus on four of the most common liquid crystal phases:  nematic liquid crystals and smectic liquid crystals in phases A, B, and C. See Figure \ref{fig 1} for a graphical representation of various phases of liquid crystals. At low temperatures, the system exists in the crystalline solid phases, spontaneously breaking all spatial translational and rotational symmetries; the corresponding EFT is therefore given by~(\ref{solid EFT no charge}) or~(\ref{charged solid eft}). As the temperature rises, symmetries are restored until at high temperatures, the system exists in fluid phase in which no symmetries other than boosts are spontaneously broken; the corresponding EFT is therefore given by~(\ref{zeroth order fluid action}) or~(\ref{zeroth order fluid action with charge}). The aim of this section will be to construct EFTs for the intermediate phases that partially break the ISO(3) group of spatial translations and rotations.

Notice that in Figure \ref{fig 1}, (a) spontaneously breaks all translations and rotations, corresponding to crystalline solid phase. Parenthetically, the profile of smectic liquid crystal in phase B appears almost identical to that of the solid; however in phase B, the horizontal layers of molecules are allowed to slide past each other. Phases (b) and (c) both spontaneously break translations in the vertical direction, but not in the horizontal directions; therefore both exist in smectic liquid crystal phase. The difference between these two states of matter is that (b) spontaneously breaks {\it all} spatial rotations whereas (c) does not break rotations about the vertical direction; thus (b) depicts phase C and (c) depicts phase A.  Phase (d) does not spontaneously break spatial translations in any direction, but does spontaneously break rotations about the horizontal axes, meaning that it exists in nematic liquid crystal phase. Finally, (e) does not spontaneously break any spatial translations or rotations and is therefore in isotropic fluid phase.


\subsection{Nematic liquid crystals}\label{Nematic}

Liquid crystals in nematic phase are composed of oblong molecules that, like the molecules of an ordinary fluid, bounce around chaotically and therefore cannot form a lattice structure; however, on average the long axes of the molecules align, thereby breaking isotropy. Because no lattice structure can form, spacetime translations remain unbroken, but the aligned long axes of the molecules spontaneously break rotations. Non-relativistically, the order parameter associated with broken rotations is a unit vector $\vec n$, pointing parallel to the long axes of the molecules. As a result the $SO(3)$ symmetry group of spatial rotations is broken to $SO(2)$~\cite{Mermin}. However, for any choice of $\vec n$, the long axes of the molecules could be equally well specified by $-\vec n$. Therefore, it is common to use the order parameter  $Q_{ij}\equiv n_i n_j-\frac{1}{3} \delta_{ij},$ which is invariant under the $\mathbb Z_2$ symmetry $\vec n\to -\vec n$. 
 To extend this order parameter to the relativistic case, it is natural to define $\cQ_{\m\n} \equiv Q_{ij} \delta^i_\m \delta^j_\n$. We therefore see that boosts are spontaneously broken as well. 
Then, relativistically, the SSB pattern of the Poincaré group is 
\begin{equation} \text{ISO(1,3)} \to \mathbb R^4 \times SO(2),  \end{equation} 
where $\mathbb R^4$ represents spacetime translations. We also have the unbroken discrete $\mathbb Z_2$ symmetry associated with $\vec n\to-\vec n$. 
Without loss of generality choose $\vev{\vec n} =\hat z $ and let indices $A,B = 1,2$.\footnote{The indices $A,B$ are not to be confused with the indices indicating unbroken symmetry generators that we used in previous sections. } Then 
$P_\m$ and $J_{3}$ 
are the unbroken generators and
$J_{ A}$ and $ K_i$
are the broken generators. It will turn out that we need to use two copies of the fields to construct the leading-order action for this phase of matter. Parameterize the most general group elements by 
\begin{equation} g_s(\phi) = e^{i X_s^\m (\phi)P_\m}  e^{i \eta_s^i(\phi) K_i}  e^{i \theta_s^{A}(\phi) J_{A}}e^{{i} \theta_s^{3}(\phi) J_{3}}.  \end{equation} 
The resulting Maurer-Cartan one-forms are
\begin{equation} g_s^{-1}\partial_M g_s = i E^\m_{sM}\cur{ P_\m +\nabla_\m \theta_s^{ A} J_{A} + \nabla_{\m} \eta_s^i K_i  } + {i} \Omega_{sM}^{3} J_{3},  \end{equation}
such that 
\begin{equation}\begin{split}   E^\m_{sM} & = \partial_M X_s^\n {[\Lambda_s R_s]_\n}^\m, 
\\  \nabla_\m \theta_s^{A} & =(E_s^{-1})_\m^M \epsilon^{AB 3} [R_s^{-1}\Lambda_s^{-1} \partial_M (\Lambda_s R_s)]^{ B3},
\\ \nabla_\m \eta_s^i & = (E_s^{-1})^M_\mu [ \Lambda_s^{-1}\partial_M \Lambda_s  ]^{0j} R_s^{ji},
\\ \Omega_{sM}^{3} & =  [R_s^{-1}\Lambda_s^{-1}\partial_M (\Lambda_s R_s) ]^{12} ,
\end{split}\end{equation}
where ${\Lambda_s^\m}_\n \equiv [e^{{i} \eta_s^{i} K_{i}}{]^\m}_\n$ and $R_s^{ij}\equiv [e^{i \theta_s ^{A} J_{A}}e^{i \theta_s^3 J_3} ]^{ij}$. 
 We can impose the thermal inverse Higgs constraints 
\begin{equation} 0= E_{s0}^i = \frac{\partial X_s^\m}{\partial \phi^0} {\Lambda_{s\m}}^j R_s^{ji}, \end{equation}
which reduce to $\partial_0 X_s^\m {\Lambda_{s\m}}^i =0$ because $R_s^{ij}$ are invertible. Then, as in the ordinary fluid case, we have
\begin{equation} \frac{\eta_s^i}{\eta_s} \tanh \eta_s = -\frac{\partial_0 X_s^i}{\partial_0 X_s^t},  \end{equation}
where $\eta_s \equiv \sqrt{\eta_s^i\eta_s^i}$. Once again, $E_{s0}^t = \sqrt{-G_{s00}}$. Additionally, we can impose the unbroken inverse Higgs constraint, which removes the rotation Goldstone $\theta_s^3$. Define the vector 
 \begin{equation}\label{Aa vector} A_a^{k}\equiv \frac{1}{2} \epsilon^{ijk}(E_1)_M^i (E_2^{-1})^M_j .\end{equation}
 The unbroken inverse Higgs constraint is then $A_{a}^3=0$. At this order in the derivative expansion it is not necessary to solve this inverse Higgs constraint, so in the interest of brevity we will not. 
 
 In the case of ordinary fluids, we had the additional unbroken inverse Higgs constraints $A_{a}^A=0$. But now $J_A$ are broken so we may not impose such constraints. We therefore have a covariant building-block that mixes $1$-and $2$-fields, namely $A_{a}^{A}$. 
 Next, define $\nabla \theta_r^A$ implicitly as follows. Under a dynamical KMS transformation, we have\footnote{Notice that $\nabla \theta_r^A$ involves a time-like derivative of $\theta_r^A$.}
\begin{equation} A_a^A\to \Theta A_a^A +i\Theta \beta_0 \nabla \theta_r^A .\end{equation}
The full set of covariant building-blocks at this order in derivatives is
\begin{equation} \begin{split}A_a^A,~~~~~~~~~~\nabla \theta_{r}^{A},~~~~~~~~~~ \nabla_B \theta_{s}^{A},~~~~~~~~~~\nabla_3 \theta_{s}^{A}, ~~~~~~~~~~ G_{s00},
\end{split}  \end{equation} 
where $G_{s00}\equiv E^\m_{s0}\eta_{\m\n} E^\n_{s0}$. Thus, at leading order in the derivative expansion, the effective action is constructed from SO(2)$\times \mathbb Z_2$-invariant combinations of the above terms. 
The contribution to the non-equilibrium effective action that does {\it not} contain mixtures of 1-and 2-fields is $S_1-S_2$, where
\begin{equation}\begin{split}S_s = \int d^4 \phi \sqrt{-G_s}~\frac{1}{2} \Big[ 2 P(G_{s00})   
+ \cK_1(G_{s00}) (\nabla^{[A}\theta_s^{B]})^2 + \cK_2(G_{s00}) (\nabla_A \theta_s^A)^2 \\+\cK_3(G_{s00}) (\nabla_3\theta_s^A)^2   \Big] .\end{split} \end{equation}
Notice that we could have included a term involving $(\nabla_t \theta_s^A)^2$, but we will see that the nematic degrees of freedom are diffusive, so we will consider $\partial _0\theta_s^A$ to count at the same order as $\partial_i^2\theta_s^A$ in the derivative expansion.
In the $r,a$-basis, the contribution to the non-equilibrium effective action that contains mixtures of 1-and 2-fields is
\begin{equation} I_\text{mix} = \int d^4 \phi\sqrt{-G_r} \squ{ -M(G_{r00}) \nabla\theta_r^A A_a^A + \frac{i}{\beta_0}M(G_{r00}) (A_a^A)^2 },\end{equation}
where $\beta_0$ is the equilibrium inverse temperature and the relationship between the coefficients of the two terms is fixed by the (classical) dynamical KMS symmetry.
 Then, up to higher-order corrections, the full non-equilibrium effective action is given by 
\begin{equation} \label{nematic eft} I_\text{EFT} = S_\text{1}-S_\text{2} + I_\text{mix}.\end{equation}

To compare the above action built with the non-equilibrium coset construction to more standard results, let the unit vector $n_s^i\equiv R_s^{i 3}$ indicate in which direction the elongated molecules of the nematic are oriented. Then, working in the classical, non-relativistic limit, we have
\begin{equation}\begin{split} \nabla \theta_r^A A_a^A =\bigg [\dot n_r^i-\frac{1}{2} \big(n_r^j\partial_j \dot X_r^i-\partial^i \dot X_r^j n_r^j\big) \bigg ] \bigg [ n_a^i-\frac{1}{2} \big(n_r^k\partial_k  X_a^i-\partial^i  X_a^k n_r^k\big) \bigg ]\end{split}\end{equation} 
and
\begin{equation}\begin{split} (\nabla^{[A}\theta_s^{B]})^2 =(\vec\nabla\cdot \vec n_s)^2,~~~~~(\nabla_A \theta_s^A)^2=(\vec n_s\cdot (\vec \nabla\times \vec n_s))^2 ,~~~~~(\nabla_3\theta_s^A)^2=(\vec n_s\times (\vec\nabla \times \vec n_s))^2,
 \end{split} \end{equation}
 where the dot ($\dot{~}$) indicates differentiation with respect to $\phi^0$ and hence is the usual material derivative when acting on $\vec n_r$. 
Note that our EFT is invariant under $\vec n_s\to -\vec n_s$. For simplicity, assume that the nematic and hydrodynamic modes decouple so we can focus just on the nematic degrees of freedom. Then, the equations of motion for the rotation Goldstones can be written in the form
\begin{equation}\dot{ \vec n} = -\frac{1}{M} \frac{\delta \cF}{\delta \vec n},~~~~~~~~~~\cF \equiv 
\frac{1}{2} \int d^3 x \big[ \cK_1 (\vec\nabla\cdot \vec n)^2 +\cK_2 (\vec n\cdot (\vec \nabla\times \vec n))^2 +\cK_3 (\vec n\times (\vec\nabla \times \vec n))^2 \big],\end{equation} 
where it is understood that $M$ and $\cK_i$ for $i=1,2,3$ are evaluated on the equilibrium value of $G_{00}$. We therefore see that $\frac{1}{M}\cF$ can be interpreted as the relaxation rate times the free energy, and $\cK_1$, $\cK_2$, and $\cK_3$ as proportional to the standard splay, twist, and bend respectively~\cite{Forster}. We have thus recovered the standard nematic equations of motion.

Notice that, linearizing in $\theta^A_r$, the dispersion relation for the nematic degrees of freedom is of the form $\omega\propto ik^2$, indicating diffusion, which is an intrinsically dissipative effect. It is worth noting that this dissipation owes entirely to the fact that the action  cannot be factorized according to~(\ref{factorized eft}) because of the inclusion of the term $I_\text{mix}$. Further, the necessity of including $I_\text{mix}$---and therefore the resulting diffusive dispersion relation---is an automatic consequence of the commutation relations of the Poincaré algebra; we did not need to assume diffusive behavior {\it a priori}. 

Finally, we often expect particle number to be conserved, requiring the inclusion of an additional unbroken $U(1)$ charge $Q$. Let $\pi$ be the corresponding unbroken Goldstone. Then just as in \S \ref{Fluids} and \S\ref{Solids}, we have the additional invariant building-block 
\begin{equation}\label{Q building-block} B_0\equiv \frac{\partial \psi}{\partial\phi^0},~~~~~~~~~~ \psi(\phi)\equiv \mu X^0(\phi)+\pi(\phi).\end{equation} 
As a result, the effective action is identical to (\ref{nematic eft}) except $P$, $M$, and $\cK_i$ for $i=1,2,3$ are now functions of both $G_{00}$ and $B_0$. 

\subsection{Smectic liquid crystals}

Smectic liquid crystals are composed of oblong molecules that form layers such that along one spatial direction, the liquid crystal has a periodic structure. Without loss of generality,  take $P_3$ to be the spontaneously broken translation generator orthogonal to the layers. To ensure that some notion of unbroken translations along the  $\hat z$ direction exists, it is necessary to introduce an internal translation symmetry generated by $Q_3$ that is spontaneously broken such that the diagonal subgroup generated by $\bar P_3\equiv P_3+Q_3$ is unbroken. Throughout this section, we will use $A,B=1,2$ to indicate spatial indices perpendicular to the $\hat z$ direction. 

\subsubsection{Phase A} \label{Smectic A}

In phase A, the long axes of the molecules are, on the average, aligned with the $\hat z$ direction. Thus,
$ \bar P_\m = P_\m+\delta^3_\m Q_3$, and $J_3$
are the unbroken generators and
$ J_A$, $K_i$, and $Q_3$
are the broken generators. The most general group element is\footnote{We use a somewhat non-standard parameterization of $g$ so that the inverse Higgs constraints are easier to solve.} 
\begin{equation} g(\phi) = e^{i X^\m (\phi)\bar P_\m} e^{i \pi^3(\phi) Q_3}  e^{i\eta^3(\phi) K_3+i \theta^{A}(\phi) J_{A}+{i} \theta^{3}(\phi) J_{3}} e^{i \eta^A(\phi) K_A} .  \end{equation} 
The resulting Maurer-Cartan one-form is
\begin{equation} g^{-1}\partial_M g = i E^\m_M\cur{ \bar P_\m +\nabla_\m \pi^3 Q_3+\nabla_\m \theta^{A} J_{A} + \nabla_\m \eta^i K_i  } + {i} \Omega_M^{3} J_{3},  \end{equation}
such that
\begin{equation}\begin{split}   E^\m_M & = \partial_M X^\n {[L \Lambda ]_\n}^\m, 
\\ \nabla_\mu \pi^3 & = (E^{-1})^M_\m \partial_M\psi^3 - \delta_\m^3,
\\  \nabla_\m \theta^{A} & = (E^{-1})^M_\m \epsilon^{AB 3} [\Lambda^{-1} L^{-1} \partial_M (L \Lambda )]^{ B3}
\\  \nabla_\m \eta^{i} & = (E^{-1})^M_\m  [\Lambda^{-1} L^{-1} \partial_M (L \Lambda )]^{0 i}
\\ \Omega_M^{3} & =  [\Lambda ^{-1} L^{-1}\partial_M {(L \Lambda)}]^{12} ,
\end{split}\end{equation}
where $\psi^3 \equiv X^3+\pi^3$ and we have ${\Lambda^\m}_\n \equiv [e^{{i} \eta^{A} K_{A}}{]^\m}_\n$ and ${L^\m}_\n \equiv [ e^{i\eta^3K_3+i \theta ^{A} J_{A} + i \theta^3 J_3} {]^\m}_\n$. Begin by imposing unbroken inverse Higgs constraints; as in the previous subsection, they serve to remove the rotation Goldstone $\theta^3$. And once again, solving these unbroken inverse Higgs constraints is not necessary at this order in the derivative expansion. Next, impose ordinary inverse Higgs constraints $\nabla_t\pi^3= \nabla_A \pi^3 = 0$, which remove $\theta^A$ and $\eta^3$ and tell us that 
\begin{equation} \nabla_3 \pi^3=\sqrt{G^{MN}\partial_M\psi^3\partial_N\psi^3}-1,\end{equation}
 yielding the first building-block
\begin{equation} Y\equiv G^{MN}\partial_M\psi^3\partial_N\psi^3, \end{equation}
where the metric is given as usual by $G_{MN}=E^\m_M\eta_{\m\n}E^\n_N$. 
Now impose thermal inverse Higgs constraints to remove $\eta^A$ given by
\begin{equation} 0= E_0^A = \frac{\partial X^\m}{\partial \phi^0} {[L \Lambda] _\m} ^A . \end{equation}
Then, we have
\begin{equation} \frac{\eta^A}{\eta_\perp} \tanh \eta_\perp = -\frac{\partial_0 X^\m {L_\m}^A}{\partial_0 X^\m {L_\m}^t},  \end{equation}
where $\eta_\perp \equiv \sqrt{\eta^A\eta^A}$. This allows us to construct our final two building-blocks $G_{00}=E^\m_0 \eta_{\m\n} E^\n_0$ and  $E^\m_0 \nabla_\m \pi^3$, the second of which is just
\begin{equation} Z\equiv \frac{\partial \psi^3}{\partial \phi^0}.  \end{equation} 
Thus, the leading-order action for smectic liquid crystals in phase-A is
\begin{equation}\label{smectic A eft}S_\text{EFT} = \int d^4 \phi \sqrt{-G} ~P(Y,Z,G_{00}) . \end{equation} 
To compare with known results, let us now expand this EFT to quadratic order in the fields. For simplicity, we will neglect the fluid degrees of freedom. Then, transforming to the physical spacetime, the quadratic Lagrangian describing the behavior of the smectic mode is 
\begin{equation}\begin{split}\cL^{(2)} = -\frac{1}{2} \Big[ M_0 (\dot \pi ^3)^2 -M_1 (\partial_A\pi^3)^2- M_3 (\partial_3 \pi^3)^2\Big],\end{split} \end{equation}
for constants $M_0$, $M_1$, and $M_3$. 
The resulting equations of motion,
\begin{equation} \ddot\pi^3 = \frac{M_1}{M_0} \partial_A^2 \pi^3 +  \frac{M_3}{M_0} \partial_3^2 \pi^3, \end{equation} 
 agree with the results of~\cite{Smectic A paper,unified hydro}. For readers interested in the full equations of motion, they can be obtained by setting $\psi^1=\psi^2=0$ in the solid equations of motion in \S\ref{Solids}. 

Finally, if particle number is conserved, we include the additional building-block~(\ref{Q building-block}),  meaning that $P$ in (\ref{smectic A eft}) may now depend on $B_0$.

\subsubsection{Phase B}

Smectic liquid crystals in phase B, like in phase A, consist of vertically stacked layers that can slide past each other. In phase A, the molecules within a given layer are able to move around freely without forming a lattice structure. In phase B, however, the molecules in a given layer are locked in place. If we take the $\hat z$ direction to be perpendicular to the stacked layers, then phase B is essentially a solid that cannot sustain uniform $ x$-$ z$ and $ y$-$ z$ shears~\cite{unified hydro}. As a result, the effective action is almost identical to (\ref{solid EFT no charge})---or if it contains an unbroken $U(1)$ charge it is almost identical to (\ref{charged solid eft}). The only difference is that to allow the layers to slide past each other, we must impose the symmetries
\begin{equation} \psi^A\to\psi^A+f^A(\psi^3), \end{equation} 
where $A=1,2$ and $f^A$ is an arbitrary function of $\psi^3$. Physically, these symmetries indicate that we can translate each layer in the $ x$-$ y$ plane independently without changing the macroscopic state of the system. At the level of the EFT, these additional symmetries mean that the effective action can only depend on $Y^{ij}$ in the combinations
\begin{equation}b\equiv \det Y^{ij}, ~~~~~~~~~~ b_1\equiv Y^{11}Y^{33}-(Y^{13})^2,~~~~~~~~~~b_2\equiv Y^{22}Y^{33}-(Y^{23})^2,~~~~~~~~~~Y^{33},  \end{equation}
and it can only depend on the $i=3$ component of $Z^i$. Finally, the equations of motion are just a special case of the solid equations of motion given in \S\ref{Solids}.

\subsubsection{Phase C}

Phase C is much like phase A except now the long axes of the molecules do not on average align with the $\hat z$ direction, meaning that $J_3$ is now spontaneously broken. Thus the only difference between phases A and C is that phase C has a Goldstone associated with the broken generator $J_3$ denoted by $\theta^3$. Further, this Goldstone cannot be removed with inverse Higgs-type constraints. 
 
Going through a similar procedure to the one in \S\ref{Smectic A}, we find that the effective action has the same building-blocks as (\ref{smectic A eft}) with five invariant additions. First, we have $\nabla_\mu\theta^3$,
where $\mu=0,1,2,3$ are now merely labels and do not need to be contracted in any particular way since the entire Lorentz group is spontaneously broken. Additionally, we have $A_a^3$, which is the $\hat z$-component of (\ref{Aa vector}). The addition of the building-block $A_a^3$, which mixes 1-and 2-fields, means that the non-equilibrium effective action cannot factorize into the difference of ordinary actions. 

The contribution to the non-equilibrium effective action that does {\it not} contain mixtures of 1-and 2-fields is $S_1-S_2$, where
\begin{equation}S_s = \int d^4 \phi \sqrt{-G_s} ~\squ{P(Y_s,Z_s,G_{s00}) +\frac{1}{2} M^{ij}(Y_s,Z_s,G_{s00}) \nabla_i\theta_s^3\nabla_j\theta_s^3} , \end{equation} 
such that repeated indices $i,j$ are summed over and $M^{ij}$ is symmetric under exchange of $i$ and $j$.\footnote{The sum over $i$ and $j$ is purely for notational convenience as $i$ and $j$ are merely labels.} Just as in the case of nematic phase, we will see that the rotation Goldstone, $\theta^3$ is diffusive, so we will consider $\partial_0\theta^3$ and $\partial_i^2\theta^3$ to be the same order in the derivative expansion.  

Define $\nabla \theta_r^3$ implicitly as follows. Under a dynamical KMS transformation, we have
\begin{equation} A_a^3\to \Theta A_a^3 +i\Theta \beta_0 \nabla \theta_r^3 .\end{equation}
In the $r,a$-basis, the contribution to the non-equilibrium effective action that contains mixtures of 1-and 2-fields is
\begin{equation} I_\text{mix} = \int d^4 \phi\sqrt{-G_r} \squ{ -M_0(G_{r00}) \nabla\theta_r^3 A_a^3 + \frac{i}{\beta_0}M_0(G_{r00}) (A_a^3)^2 }.\end{equation}
As a result, up to higher-order corrections, the full non-equilibrium effective action is
\begin{equation} \label{nonequ EFT phase C}I_\text{EFT} = S_\text{1}-S_\text{2} +I_\text{mix}.\end{equation}

If particle number is conserved, we have the additional building-block~(\ref{Q building-block}). Then the non-equilibrium effective action is identical to (\ref{nonequ EFT phase C}), except the coefficient functions $P$, $M_{ij}$, and $M_{0}$ may now depend on $B_0$.

The full equations of motion for smectic C are quite complicated and therefore not terribly enlightening. Thus, for the sake of simplicity assume that the dynamics of $\theta^3$ decouple from those of $X^\mu$ and $\pi^3$. Then, the equations of motion for $X^\mu$ and $\pi^3$ are identical to those of the smectic A phase. And the linearized equation of motion for $\theta^3$ is, ignoring the complications of hydrodynamical fluctuations,
\begin{equation}\dot\theta^3 = \frac{M^{ij}}{M_0}   \partial_i \partial_j \theta^3. \end{equation} 
Thus, smectic C phase looks just like smectic A phase with the addition of a diffusive mode $\theta^3$, agreeing with standard results~\cite{unified hydro}. 

It is worth noting that the presence of the term $I_\text{mix}$, as in the nematic phase,  leads to a diffusive dispersion relation and hence dissipation. Thus nematic and smectic C phases of liquid crystal are the only two states of matter considered in this paper that exhibit dissipation at leading order in the derivative expansion.


\section{Conclusion}

In this paper, we defined a systematic coset construction of non-equilibrium EFTs and used it to formulate both known and heretofore  unknown non-equilibrium EFTs for condensed matter systems. We postulated that IR dynamics of thermal systems out of equilibrium can be characterized by Goldstones almost as if every symmetry of the system were spontaneously broken. However, the Goldstones corresponding to spontaneously broken symmetry generators behave rather differently than those corresponding to unbroken symmetry generators. In particular, the unbroken Goldstones possess infinitely many gauge symmetries, whereas the broken Goldstones, like ordinary Goldstones at zero temperature, have no such gauge symmetries. 

The approach of~\cite{coset} was to treat these infinitely many symmetries as if they were true symmetries of the underlying theory. As a result, each of these symmetries required its own generator and Goldstone to parameterize the coset of broken symmetries.\footnote{Since~\cite{coset}  only constructed actions to leading order in the derivative expansion, it was not actually necessary to introduce all of the infinitely many symmetry generators. However, extending to arbitrarily higher orders using their method would require the addition of infinitely many symmetry generators and Goldstones; see~\cite{Joyce} for a general discussion of how gauge symmetries can be treated as genuine symmetries in the coset construction.} By contrast, our approach was to treat these additional symmetries as gauge redundancies in the style of~\cite{Wheel}, thereby circumventing the need to introduce infinitely many symmetry generators and Goldstones. Further, the coset construction presented in this paper allows one to formulate non-equilibrium actions in full generality, whereas previous attempts at the coset construction for systems with spontaneously broken spacetime symmetries can only be used to formulate ordinary actions that are incapable of accounting for statistical fluctuations and dissipation.  

The non-equilibrium coset construction admits generalizations and applications in many directions. First, in the interest of simplicity, the EFTs in this paper were computed to lowest order in the derivative expansion. As a result, all but the EFTs for nematic and smectic C phases of liquid crystal admit nothing of statistical fluctuations and dissipation. We leave it as future work to `turn the crank' and extend the actions presented here to higher orders in the derivative expansion. Second, the ordinary coset construction presented in~\cite{Weinberg} can be used to construct pseudo-Goldstone EFTs for spontaneously broken approximate symmetries. It will be of interest to extend the non-equilibrium coset construction to cases for which approximate symmetries exist. By the philosophy of this paper, there ought to be associated pseudo-Goldstones whether or not the approximate symmetries are spontaneously broken. 
Third, the EFTs we constructed for solids and smectic liquid crystal phases exhibit second sound. While second sound can exist in such phases of matter, it is exceedingly rare~\cite{solid second sound ref}. We leave it as future work to construct EFTs for these states of matter without second sound. 
Fourth, some condensed matter systems have excitations that survive over long distances and extended time scales but that cannot be interpreted as Goldstone modes; for example, such modes exist near critical points. Thus, extending the non-equilibrium coset construction to allow for couplings between Goldstone and non-Goldstone modes is of significant interest. 
Finally, the hydrodynamic degrees of freedom in the non-equilibrium EFTs of~\cite{H. Liu} do not necessitate a long-wavelength expansion. However our non-equilibrium coset construction---like the ordinary coset construction---is designed to generate covariant terms for an action in a derivative expansion. Extending our coset construction to allow for non-local terms would allow one to formulate actions with much broader applicability, e.g. actions describing low-temperature systems. 

\bigskip

\noindent {\bf Acknowledgments:} I am pleased to dedicate this paper to Dorian Goldfeld, Professor of Mathematics at Columbia University. I would like to thank Alberto Nicolis and Lam Hui for their wonderful mentorship as well as Austin Joyce and Noah Bittermann for many insightful conversations. This work was partially supported by the US Department of Energy grant $\text{DE-SC0011941}$.

\appendix

\section{Emergent gauge symmetries}\label{Symmetry origin}

In this section, we offer a somewhat pragmatic explanation for why we expect non-equilibrium effective actions to possess the gauge symmetries~(\ref{partial gauge 0}-\ref{partial gauge 1}). It is pragmatic in the sense that it should convince the reader that these symmetries are necessary in ordinary situations, but it is by no means a derivation of the symmetries from first principles. 

In the ordinary construction of the hydrodynamic equations of motion it is usually supposed that the system exists in `local equilibrium.' This essentially means that, ignoring broken Goldstones, the state of the system can be specified by the local inverse temperature four-vector $\beta^\m(x)$ as well as the chemical potentials $\mu^A(x)$ corresponding to unbroken symmetries $T_A$. From the EFT perspective, we posit that the fluid manifold is a space on which the fluid four-vector is fixed, namely $\beta^M=(\beta_0,0,0,0)^M$, where $\beta_0$ is the equilibrium inverse temperature. Then, in the physical spacetime, we can define the inverse-temperature four-vectors---one for each leg of the SK contour---via the push-forwards
\begin{equation}\label{beta}\beta_s^\m(x)\equiv \beta^M\frac{\partial X_s^\m}{\partial\phi^M} = \beta_0 \frac{\partial X_s^\m}{\partial\phi^M}. \end{equation}
Additionally, we can identify the chemical potential with\footnote{It may not be immediately obvious that this combination of fields should be identified with the chemical potential. For some insight into why this should be so, see the discussion surrounding (\ref{gauge field expression}-\ref{zeroth order fluid action with charge}) and \cite{H. Liu 2,MHD}.}
\begin{equation}\label{mu}\mu_s^A(x) = \mu^A_0 + \frac{\cB_{s0}^A}{\sqrt{-G_{00}}}, \end{equation} 
where $\cB_{s0}^A$ is given in~(\ref{MC form}) and  $\m_0^A$ are the equilibrium chemical potentials of the unbroken symmetries. Working in the $r,a$-basis of (\ref{r a variables}), the $r$-type fields correspond to classical field configurations, whereas the $a$-type fields describe quantum and statistical fluctuations. As a result, we ought to identify the physical thermodynamic quantities $\beta^\m(x)$ and $\mu^A(x)$ with
\begin{equation}\label{physical local parameters}\begin{split}\beta^\m(x)& =\beta^\m_r(x)\equiv\frac{1}{2}\squ{\beta^\m_1(x)+\beta^\m_2(x)},
\\ \mu^A(x)& =\mu^A_r(x)\equiv\frac{1}{2}\squ{\mu^A_1(x)+\mu^A_2(x)}.
\end{split}\end{equation}
If we wish to reproduce ordinary hydrodynamics, then we must require that the equations of motion only depend on the unbroken Goldstone fields as they appear in (\ref{physical local parameters}).
To ensure that this is the case, we must require that the effective action be invariant under the symmetries~(\ref{partial gauge 0}-\ref{partial gauge 1}). Further, since we expect that the (classical) state of the system is specified by $\beta_r^\m(x)$ and $\mu_r^A(x)$, these emergent symmetries cannot change the state of the system and therefore ought to be considered as mere gauge redundancies. 

It is worth mentioning that beyond these heuristic arguments, there has been some success in deriving chemical shift symmetries for unbroken $U(1)$ Goldstones using the AdS/CFT duality from first principles~\cite{Holography chemical shift 1,Holography chemical shift 2}.

\section{Stückelberg tricks  and the Maurer-Cartan form}\label{Stückelberg}

We claimed in \S\ref{non-eq coset section} that the correct way to construct covariant building-blocks for non-equilibrium effective actions was to construct two distinct Maurer-Cartan one-forms~(\ref{MC form})---one for each leg of the SK contour---that transform under a single copy of the global symmetry group $\cG$. In this section, we will use a Stückelberg trick inspired by~\cite{H. Liu} to demonstrate that using two copies of the Maurer-Cartan form is the correct approach. 

We begin by introducing sources for the Noether currents corresponding to each symmetry generator of $\cG$. This amounts to introducing external gauge fields. Let 
\begin{equation}\begin{split} \bar P_m & =\text{unbroken translations},
\\ T_A & = \text{other unbroken generators},
\\ \tau_\alpha & =\text{broken generators},
 \end{split}\end{equation}
 be the generators of $\cG$, where we now use $m,n=0,1,2,3$ to denote Lorentz indices and $\m,\n$ to denote physical spacetime coordinate indices.\footnote{It is now necessary to distinguish between Lorentz indices $m,n$ and physical spacetime coordinate indices $\m,\n$ because the Stückelberg trick requires that we gauge all symmetries including Lorentz. } As before, let $\cH$ be the set of unbroken symmetries and $\cH_0\subset \cH$ the subgroup generated by $T_A$ alone.  The external sources are as follows:
\begin{itemize}
\item Let $\varepsilon_{s\m}^m(x)$ be the vierbeins, which can be thought of as the gauge fields corresponding to unbroken translations $\bar P_m$.  The metric tensors are then given by $g_{s\m\n}(x) = \varepsilon_{s\m}^m(x)\eta_{mn}  \varepsilon_{s\n}^n(x)$. 
\item Let $\cA_{s\m}(x)\equiv \cA_{s\m}^A(x) T_A$ be the gauge fields (or spin connections if the Lorentz group is involved) corresponding to the unbroken symmetries of $\cH_0\subset \cH$. 
\item Let $c_{s\m}(x) \equiv c_{s\m}^\alpha(x) \tau_\alpha $ be the gauge fields (or spin connections) corresponding to the broken symmetries. 
\end{itemize}
On each leg of the SK contour, we can combine these fields into a single object,
\begin{equation} \theta_{s\m}(x) = i \varepsilon_{s\m}^m(x)\bar P_m+ i c_{s\m}^\alpha(x) \tau_\alpha + i\cA_{s\m}^A(x) T_A, \end{equation}
where the factors of $i$ are included for later convenience. Now, letting $U(t,t';\theta_{s\m})$ for $s=1,2$ be the time evolution operator from $t'$ to $t$ in the presence of external sources $\theta_{s\m}$, the generating functional for the conserved currents is
\begin{equation}e^{W[\theta_{1\m},\theta_{2\m}]} = \text{\tr}\squ{U(+\infty,-\infty; \theta_{1\m})\rho U^\dagger(+\infty,-\infty; \theta_{2\m})}. \end{equation}
Since $\theta_{s\m}$ couple to conserved currents, $W[\theta_{1\m},\theta_{2\m}]$ must be invariant under two independent copies of the gauge symmetries~\cite{H. Liu}; that is, for gauge parameters $\zeta_1(x)$ and $\zeta_2(x)$ we have
\begin{equation}W[\theta_{1\m},\theta_{2\m}] = W[\theta_{1\m}^{\zeta_1},\theta_{2\m}^{\zeta_2}].\end{equation} 
We can therefore `integrate in' both the broken and unbroken Goldstone fields using the Stückelberg trick. In particular, define
\begin{equation}\Theta_{sM}(\phi)\equiv \gamma_s^{-1}\cdot \squ{ \theta_{s\m}(X_s(\phi)) \frac{\partial X_s^\m}{\partial\phi^M}+\frac{\partial}{\partial\phi^M}}\gamma_s,~~~~~~~~~~\gamma_s(\phi) \equiv e^{i\pi^\alpha_s(\phi)\tau_\alpha}e^{i\epsilon_s^A(\phi) T_A}, \end{equation}
where $\phi^M$ for $M=0,1,2,3$ are the fluid worldvolume coordinates. 
Now we can implicitly define the non-equilibrium effective action by 
\begin{equation}e^{W[\theta_{1\m},\theta_{2\m}]} \equiv \int \cD X_s^\m\cD \pi_s^\alpha\cD \epsilon_s^A ~e^{i I_\text{EFT}[\Theta_{1M},\Theta_{2M}]}.\end{equation}
Notice that if we remove the external source fields by fixing $\varepsilon_{s\m}^m(x)=\delta_\m^m$ and $\cA_{s\m}^A=c_{s\m}^\alpha=0$, we find that $\Theta_{sM}$ are nothing other than the Maurer-Cartan one-forms~(\ref{MC form}) defined on each leg of the SK contour. Since $\varepsilon_{s\m}^m(x)=\delta_\m^m$ we can identify the Lorentz indices $m,n$ with the physical spacetime coordinate indices $\m,\n$, allowing us to use $\bar P_\m$ to refer to the unbroken translation generators as we did in~(\ref{the symmetry generators}). Moreover because these fields live on the SK contour, their values must match up in the infinite future, meaning that even though there were two copies of the gauge fields and gauge symmetries, there is only one copy of the global symmetry group $\cG$.

\section{Explicit inverse Higgs computations}\label{explicit IH computations}

Since the thermal and unbroken inverse Higgs constraints are new, it may be helpful to see how they can be solved explicitly. This appendix explains how to algebraically solve the thermal and unbroken inverse Higgs constraints in the case of ordinary fluids, thereby removing the broken boost Goldstones $\eta^i$ and the unbroken rotations Goldstones $\theta^i$, respectively. 
But before we solve these new inverse Higgs constraints at full nonlinear order, it is first helpful to understand the motivations for imposing them. To do this, we will investigate these inverse Higgs constraints expanded to linear order in the fields. 

Consider the case of a fluid without conserved charge. Let $X_s^\m(\phi)=\phi^\m+\varepsilon_s^\m(\phi)$ and let the antisymmetric field $\theta_s^{\m\n}(\phi)$ represent all Lorentz Goldstones such that $\eta_s^i=\theta_s^{0i}$ and $\theta_s^i=\frac{1}{2} \epsilon^{ijk}\theta_s^{jk}$. Then at linear order in the fields, the vierbeins are given by
\begin{equation}E_{sM}^\m =\delta^\m_M+\partial_M \varepsilon_s^\m -{\theta_{sM}}^\m+\cdots,\end{equation} 
where $s=1,2$ indicates on which leg of the SK contour the fields are defined. We are interested in setting to zero components of $E_{sM}^\m$ that transform covariantly and that can give an algebraic relation between ${\theta_{sM}}^\m$ and derivatives of $\varepsilon_s^\m$.

To remove the broken boost Goldstones, notice that $E_{s0}^i$ for $i=1,2,3$ transform covariantly under all symmetries and gauge symmetries and are given by
\begin{equation} E_{s0}^i = \partial_0\varepsilon_s^i-{\theta_{s0}}^i  +\cdots. \end{equation}
Thus setting the above expression to zero gives the linearized constraint 
\begin{equation} \eta^i_s=-\partial_0 \varepsilon_s^i ,  \end{equation}
which is just the linearized version of~(\ref{thermal IH ordinary fluids...}). This allows the removal of the $\eta_s^i$-Goldstones. 

Removal of the rotational Goldstones is a bit trickier. There are no terms in a single vierbein that transform covariantly and, when set to zero, allow the removal of $\theta^i_s$. However, we have two vierbeins---one for each leg of the SK contour---thus we consider the object $E_{1M i} (E_2^{-1})^M_j$, which transforms covariantly under all symmetries and gauge symmetries. Setting the antisymmetric part to zero gives the relation
\begin{equation}0=E_{1M i} (E_2^{-1})^M_j - E_{1M j} (E_2^{-1})^M_i = \partial_j\varepsilon_{ai} -\partial_i \varepsilon_{aj} - 2 {\theta_{aji}} +\cdots,\end{equation}
where $\varepsilon_a^i \equiv \varepsilon_1^i-\varepsilon_2^i$ and $\theta_a^{ij}\equiv \theta_1^{ij}-\theta_2^{ij
}$. Solving the above equation gives the constraint
\begin{equation}\label{unbroken 1 apdx}\vec \theta_a = \vec\nabla\times\vec\varepsilon_a,  \end{equation} 
which is the linearized version of~(\ref{unbroken IH ordinary fluids...}). Then, acting with the (classical) dynamical KMS symmetry on both sides gives
\begin{equation}\label{unbroken 2 apdx}\partial_0 \vec \theta_r = \vec\nabla\times\partial_0\vec\varepsilon_r,\end{equation}
where $\varepsilon_r^i \equiv \frac{1}{2} \cur{\varepsilon_1^i+\varepsilon_2^i}$ and $\theta_r^{ij}\equiv \frac{1}{2}({\theta_1^{ij}+\theta_2^{ij}})$. Since $\vec\theta_r$ enjoys a chemical shift-type gauge symmetry, at linear order, $\vec\theta_r$ can only appear in the form $\partial_0\vec\theta_r$. Thus, the constraints~(\ref{unbroken 1 apdx}-\ref{unbroken 2 apdx}) are sufficient to remove the $\vec\theta_s$-Goldstones entirely from the (quadratic) effective action.

\subsection{Thermal inverse Higgs} \label{c1}

Recall that for ordinary fluids, the thermal inverse Higgs constraints are given by~(\ref{thermal IH ordinary fluids}). Since $R$ is invertible, these constraints give
\begin{equation} \label{thermal IH appendix} 0= \frac{\partial X^\m }{\partial \phi^0}{\Lambda_\m}^i =  \frac{\partial X^t }{\partial \phi^0}{\Lambda_t}^i +  \frac{\partial X^j }{\partial \phi^0}{\Lambda_j}^i .\end{equation} 
Let $\lambda$ be the $3 \times3$ matrix with components ${\lambda^i}_j = {\Lambda_j}^i$ and let $\vec l$ be the 3-vector with components $l^i={\Lambda_t}^i$. Then we have
\begin{equation}\lambda = (1-\hat \eta \otimes \hat \eta ) +\hat \eta\otimes\hat \eta \cosh\eta, ~~~~~~~~~~\vec l = \hat \eta \sinh\eta, \end{equation} 
where $\eta\equiv \sqrt{\vec\eta\cdot\vec\eta}$ and $\hat \eta \equiv \vec\eta/\eta$.
Now equation~(\ref{thermal IH appendix}) becomes 
\begin{equation} \frac{\partial_0\vec X}{\partial_0 X^t} = -\lambda^{-1} \cdot \vec l, \end{equation} 
which simplifies to
\begin{equation}  \frac{\partial_0 X^i}{\partial_0 X^t} = -\frac{\eta^i}{\eta} \tanh \eta ,\end{equation}
as desired. 

\subsection{Unbroken inverse Higgs}\label{c2}

Recall that for ordinary fluids, the unbroken inverse Higgs constraints are given by~(\ref{unbroken IH ordinary fluids}). Expanding out the vierbeins, we find that
\begin{equation}E_{sM}^\m = {a_{sM}}^j {R_{sj}}^i,~~~~~~~~~{a_{sM}}^\n\equiv \frac{\partial X_s^\rho}{\partial\phi^M} {\Lambda_{s\rho}}^\nu,\end{equation} 
where $R_s^{ij} =[e^{i\theta_s^i(\phi)J_i}]^{ij}$. Then, we have that 
\begin{equation} (E_s^{-1})^{M}_i  =  (R_s^{-1}{)_i}^{j}(a_s^{-1}{)_j}^M,    \end{equation}  
which gives us
\begin{equation} (E_1)_{i M} (E^{-1}_2{)^M}_j = (a_1{)_M}^l (R_1)_{l i} (R_2^{-1}{)_j}^k (a_2^{-1}{)_k}^M \equiv (R_2^{-1}{)_j}^k {\cM_k}^l (R_1)_{li}, \end{equation}
where $ {\cM_k}^l \equiv (a_2^{-1}{)_k}^M (a_1{)_M}^l    = (\Lambda_2^{-1}{)_k}^{\n} \frac{\partial X_1^\m}{\partial X_2^\n} (\Lambda_1{)_\m}^l$. Then unbroken inverse Higgs constraints (\ref{unbroken IH ordinary fluids}) yield the $3\times 3$ matrix equation
\begin{equation} R_2^{T} \cdot \cM\cdot R_1 = R_1^T \cdot \cM^T \cdot R_2,  \end{equation} 
where we have used the fact that $R_s^T=R_s^{-1}$. 
Rearranging this equation we have that $\cM^T = (R_1\cdot R_2^{-1}) \cdot \cM\cdot (R_1\cdot R_2^{-1})$, from which it is immediate that
\begin{equation} \cM^T\cdot \cM = (R_1\cdot R_2^{-1}) \cdot \cM\cdot \cM^T \cdot(R_1\cdot R_2^{-1})^T = (R_1\cdot R_2^{-1} \cdot\cM)^2. \end{equation} 
Finally, we have that $(R_1\cdot R_2^{-1})\cdot \cM = \sqrt{\cM^T\cdot \cM}$, that is,
\begin{equation}\label{final unbroken IH} R_1\cdot R_2^{-1} = \sqrt{\cM^T\cdot \cM}\cdot\cM^{-1} , \end{equation} 
as desired. 
%
Notice that the above expression, when linearized, gives (\ref{unbroken 1 apdx}), so it can be used to remove $a$-type rotational Goldstones. To remove the $r$-type rotational Goldstones beyond the linearized level requires a bit more work; we will carry out the computation in the next subsection.

\subsection{More on unbroken inverse Higgs}\label{c3}

Now we will see that the unbroken inverse Higgs constraints allow the removal of the $r$-type rotational Goldstones beyond linear order in the fields. First, define $M_a \equiv \sqrt{\cM^T\cdot \cM^{-1}}$ and suppose that under the (classical) dynamical KMS symmetries, we have
\begin{equation} \label{Ma and Mr} M_a\to\Theta M_a - i\beta \Theta \partial_0 M_r, \end{equation} 
where $\Theta$ is a time-reversing symmetry of the UV theory.\footnote{Think of (\ref{Ma and Mr}) as a definition of $M_r$.  } Then, by applying the (classical) dynamical KMS transformations to (\ref{final unbroken IH}), we have
\begin{equation}\label{unbroken IH c3}R_1 \cdot \Omega_{r0} \cdot R_2^T = \partial_0 M_r~~~\implies~~~\Omega_{r0} = R_1^T\cdot \partial_0 M_r\cdot R_2,\end{equation}
where $\Omega_{rM}\equiv \frac{1}{2}\cur{\Omega_{1M}+\Omega_{2M}}$ is the $r$-type spin connection. The above equation is our second unbroken inverse Higgs constraint and when linearized, gives~(\ref{unbroken 2 apdx}). It can be checked that in every {\it invariant} building-block, $R_s$ can appear without $\phi^0$-derivatives only in the form $R_1\cdot R_2^T$. But then we can use (\ref{final unbroken IH}) to remove this combination of rotational Goldstones. Further, if $R_s$ appears in an invariant building-block, the only other package it can come in is $\Omega_{r0}$, in which case we can use (\ref{unbroken IH c3}) to remove $\Omega_{r0}$. This expedient comes at the price of introducing more factors of $R_s$. It turns out that there are two possibilities:
\begin{itemize}
\item The additional factors of $R_s$ are contracted in such a way that they cancel completely, in which case we have no remaining factors of $R_s$.
\item The additional factors of $R_s$ appear in the form $R_1\cdot R_2^T$ in which case we can use (\ref{final unbroken IH}) to remove them. 
\end{itemize}
Therefore, no matter what, the rotational Goldstones can be entirely removed from the {\it invariant} building-blocks using (\ref{final unbroken IH}) and (\ref{unbroken IH c3}). It should be noted that the {\it covariant} building-blocks may still have factors of $R_s$ attached to them that cannot be removed with inverse Higgs-type constraints, but this is not an issue as only the invariant building-blocks are relevant for constructing EFTs.

\section{Fluids and volume-preserving diffeomorphisms}\label{V diff}

It turns out that despite appearances, the effective actions for fluids with and without charge that we constructed with cosets,~(\ref{zeroth order fluid action}) and~(\ref{zeroth order fluid action with charge}) respectively,  are equivalent to those given in~\cite{Nicolis}. Additionally, the action for finite-temperature superfluids~(\ref{coset finite T superfluid}) that we constructed with cosets turns out to be equivalent to the action presented in~\cite{Finite T superfluid}.  In this section, we will see how these seemingly different actions are equivalent. For definiteness, consider the action~(\ref{zeroth order fluid action}), which is defined on the fluid worldvolume. Perform a change of coordinates so that it is now defined on the physical spacetime; then the dynamical degrees of freedom are the fields $\phi^M(x)$ and they enjoy the gauge symmetry
\begin{equation}\label{final gauge symmetry}\phi^M(x) \to \phi^M(x) +\xi^M(\phi^I(x)), \end{equation}
for arbitrary $\xi^M$. As always, we use $M,N=0,1,2,3$ and $I,J=1,2,3$. The resulting action defined on the physical spacetime is
\begin{equation}\label{fluid eft appendix} S_\text{EFT} = \int d^4 x~ P(T), \end{equation}
where\footnote{It turns out that $T^2=-1/G_{00}$. } $T\equiv u^\m \partial_\m \phi^0$ such that 
\begin{equation} u^\m\equiv \frac{J^\m}{b},~~~~~~~~~~b\equiv \sqrt{-J^\m J_\m},~~~~~~~~~~J\equiv \star\squ{d\phi^1\wedge d \phi^2\wedge d\phi^3} . \end{equation} 
Now integrate out $\phi^0$. The equation of motion for $\phi^0$ is $0 = \partial_\m \squ{ \frac{J^\m}{b}  P'(T) }$.
Since $\partial_\m J^\m\equiv 0$ identically, this becomes
\begin{equation} 0 = J^\m \partial_\m\squ{\frac{1}{b} P'(T) }.  \end{equation}
By the definition of $J^\m$, we have that
\begin{equation} \label{phi0 eom} \frac{1}{b} P'(T) = f(\phi^I), \end{equation} 
for some arbitrary function $f(\phi^I)$. Using the  diffeomorphism gauge symmetry (\ref{final gauge symmetry}),  we can  gauge-fix $f(\phi^I)=1$. With this gauge-fixing condition,~(\ref{phi0 eom}) can be solved algebraically for $T$ as a function of $b$, that is $T=T(b)$. Plugging this solution back into the expression for the effective action~(\ref{fluid eft appendix}) gives
\begin{equation}S_\text{EFT} = \int d^4 x~F(b),\end{equation}
where $F(b)\equiv P(T(b))$. But this is precisely the action given in \cite{Nicolis}. Notice that now the action only depends on the three fields $\phi^I$, which enjoy a volume-preserving diffeomorphism gauge symmetry 
\begin{equation}\phi^I \to g^I(\phi^J),~~~~~~~~~~\det\frac{\partial g^I}{\partial \phi^J} =1. \end{equation}

By almost identical procedures, it can be shown that~(\ref{zeroth order fluid action with charge}) is equivalent to the effective action describing charged fluids given in~\cite{Nicolis} and that~(\ref{coset finite T superfluid}) is equivalent to the action describing finite-temperature superfluids given in~\cite{Finite T superfluid}.

\end{document}